\documentclass[superscriptaddress,aps,prl,twocolumn,floatfix]{revtex4}

\usepackage{nqm}
\usepackage{dcolumn}
\usepackage{doi}

\begin{document}


\title{Pinhole interference in three-dimensional fuzzy space}
\author{D.~Trinchero}
\email{dario.trinchero@pm.me}
\affiliation{Department of Physics, Stellenbosch University, Matieland 7602, South Africa}
\author{F.~G.~Scholtz}
\affiliation{Department of Physics, Stellenbosch University, Matieland 7602, South Africa}
\date{\today}

\begin{abstract}
    We investigate a quantum-to-classical transition which arises naturally within the \emph{fuzzy sphere} formalism for three-dimensional non-commutative quantum mechanics. This transition may be understood as the mechanism of decoherence, but without requiring an additional external heat bath. We focus on treating a two-pinhole interference configuration within this formalism, as it provides an illustrative toy model for which this transition is readily observed and quantified. Specifically, we demonstrate a suppression of the quantum interference effects for objects passing through the pinholes with sufficiently-high energies or numbers of constituent particles.

    Our work extends a similar treatment of the double slit experiment, presented in~\cite{pittaway2021quantum}, within the two-dimensional Moyal plane, only it addresses two key shortcomings that arise in that context. These are, firstly that the interference pattern in the Moyal plane lacks the expected reflection symmetry present in the pinhole setup, and secondly that the quantum-to-classical transition manifested in the Moyal plane occurs only at unrealistically high velocities and/or particle numbers. Both of these issues are solved in the fuzzy sphere framework.
\end{abstract}

\maketitle


\section{Introduction}\label{ch:introduction}
Non-commutative quantum mechanics (NCQM) sets ordinary quantum mechanics within a spacetime with non-commutative geometry. Such models were proposed originally to regularise field theories~\cite{snyder1947quantized} --- an approach now supplanted by renormalisation. Still, NCQM remains well motivated: non-commutativity implies a minimum length scale, the need for which in a consistent formulation of quantum mechanics and gravity is compellingly argued by Doplicher \emph{et al.}~\cite{doplicher1995quantum}. Indeed, non-commutative spacetime geometries arise from simple limits of string theories~\cite{seiberg1999string,alekseev2000brane} --- popular contenders for a quantum theory of gravity.

At this stage, both quantum mechanics~\cite{scholtz2009formulation,balachandran2004unitary} and quantum field theory~\cite{douglas2001noncommutative} have been formulated on non-commutative spaces, and many classical problems from ordinary quantum mechanics have been reexamined in these formalisms, including the spherical well~\cite{chandra2014spectrum}, and three-dimensional scattering~\cite{kriel2017scattering}.

NCQM may provide insight into another large open question in modern physics, namely the \emph{measurement problem}, which pertains to the transition from quantum to classical behaviour. Historically this has been largely a distinct problem from the unification of quantum mechanics with gravity; yet, there is reason to believe that the problems may be related (for instance, arguments by Penrose~\cite{penrose1996gravity}). The connection to NCQM in particular comes in the form of a recent investigation by Pittaway \emph{et al.}~\cite{pittaway2021quantum}, which demonstrated a continuous quantum-to-classical transition that arises naturally within the Moyal plane formalism of NCQM.

Of particular relevance to this paper is the section of \cite{pittaway2021quantum} that treats double-slit interference within the non-commutative plane. Here it was found that the non-commutativity modifies the usual interference pattern in such a way that the quantum effects are suppressed for objects passing through the slits with sufficiently-high energies or numbers of constituent particles. This is one manifestation of the remarkable quantum-to-classical transition to which we alluded; it shows that the structure of space at the smallest length scales can play a direct role in the suppression of quantum effects at larger length scales.

However, the two-dimensional interference calculation comes with a number of caveats. Firstly, the commutation relations for the Moyal plane break rotational symmetry, and as such the interference pattern fails to be symmetric under reflection about its centre. Secondly, and more crucially, if one supposes that the quantum-to-classical transition should occur at non-relativistic velocities, for a number of protons on the order of Avogadro's number, one obtains a non-commutative parameter many orders of magnitude larger than the Planck length --- much larger than expected.

We address both of these caveats by moving to three dimensions, considering now two-\emph{pinhole} interference in the \emph{fuzzy sphere} formalism for three-dimensional non-commutative space, where rotational symmetry is preserved. As with the two-dimensional case, the particular interference setup serves merely as a toy model for analysing the quantum-to-classical transition in 3D fuzzy space. We find both that the problematic reflection-asymmetry in the interference pattern is absent in three dimensions, and that quantum effects are more strongly suppressed than in two dimensions; that is, we observe suppression at realistic length scales and non-relativistic speeds.

Our paper is structured as follows. Sections~\ref{ch:formalism} and~\ref{ch:free-particle} comprise background material. In section~\ref{ch:formalism} we review the formalism of NCQM in fuzzy space; we are especially thorough with our presentation here because notation and conventions vary in the literature. In section~\ref{ch:free-particle} we summarise the non-commutative free particle solutions. In section~\ref{ch:pinhole} we calculate our main result: the probability amplitude for measuring a single particle on a screen behind a pinhole interference setup in fuzzy space. In section~\ref{ch:discussion} we discuss physical implications of the result, show that it reduces in the appropriate limit to the known commutative result, and derive how the suppression effect scales with multiple particles. Finally, a summary of the results and concluding remarks are given in section~\ref{ch:conclusion}.

\section{Formalism}\label{ch:formalism}

\subsection{Fuzzy space}\label{sec:formalism}

NCQM begins with a non-commutative coordinate operator algebra. The simplest such algebra --- that with Heisenberg-Moyal commutation relations,
\(
\left[\hat{x}_i,\hat{x}_j\right] = i\theta_{ij},
\)
--- suffices for two dimensions~\cite{scholtz2009formulation,galikova2013coulomb}, but breaks three-dimensional rotational symmetry~\footnote{
	Some problems, like the Aharonov-Bohm effect~\cite{chaichian2008gauge,chaichian2002aharonov} or Hydrogen atom~\cite{adorno2009dirac,chaichian2004non,chaichian2001hydrogen} can nevertheless be considered in 3D using the symmetry-breaking Heisenberg-Moyal commutation relations. However, they are generally still undesirable --- in particular, they lead to problematic thermodynamic behaviour~\cite{kriel2012entropy}.
}
(this is apparent because the tensor $\theta_{ij}$ is skew-symmetric and so has a vanishing eigenvalue, the associated eigenvector of which is a preferred commutative direction~\cite{chandra2014spectrum}). A better alternative~\cite{scholtz2018classical,galikova2013coulomb}, adopted here, is the $\su2$ algebra of \emph{fuzzy sphere} geometry, given by
\begin{equation}\label{eq:commutation-rels}
	\left[\hat{x}_i,\hat{x}_j\right] = 2i\lambda\, \epsilon_{ijk}\hat{x}_k,
\end{equation}
where $\lambda$ is a constant with dimensions of length, called the \emph{non-commutative parameter}.

We represent this algebra concretely using the Jordan-Schwinger representation, whereby $\su2$ elements act on a two-boson-mode Fock space,
\[
	\Hc \defeq \spn\left\{\ket{n_1,n_2}\equiv
	\frac{(a_1^\dagger)^{n_1}(a_2^\dagger)^{n_2}}{\sqrt{n_1!n_2!}}\ket{0}
	\ \middle|\ n_1,n_2\in\mathbb{N}\right\},
\]
via the ``Jordan map''~\cite{schwinger1952angular},
\begin{equation}\label{eq:coord-operators}
	\hat{x}_i \mapsto \lambda\, a_\mu^\dagger\sigma^{i}_{\mu\nu}a_\nu.
\end{equation}
Here $a_\mu^\dagger$ and $a_\mu$ (for $\mu=1,2$) are the standard creation and annihilation operators respectively, and
\(
\ket{0}\equiv\ket{0,0}\in\Hc
\)
is the vacuum state. As an $\su2$-representation, $\Hc$ decomposes into irreps,
\begin{equation}\label{eq:Hc-irreps}
	\begin{split}
		\Hc &= \bigoplus_{n\in\mathbb{N}}\mathcal{F}_n, \quad\text{where}\\
		\mathcal{F}_n &\defeq \spn\left\{\ket{n_1,n_2}\in\Hc
		\ \middle|\ n_1+n_2=n\right\},
	\end{split}
\end{equation}
which are just as well considered irreps of $\SU2$, as $\SU2$ is simply-connected~\cite{hall2003lie}. These irreps are indexed by the $\su2$ Casimir operator, $\hat{r}^2=\hat{x}_i\hat{x}_i$ which can be rewritten in terms of the boson number operator,
$\hat{n}=a_\mu^\dagger a_\mu$, as
\(
\hat{r}^2 = \lambda^2 \hat{n} (\hat{n} + 2).
\)
Its square root (to leading order in $\lambda$) has the right dimensions for a measure of \emph{radius}, given by
\begin{equation}\label{eq:r}
	\hat{r} = \lambda(\hat{n}+1).
\end{equation}

Now $\Hc$ contains each $\su2$-irrep, and hence so too each quantised radius, exactly once. This motivates our choice of the Jordan-Schwinger representation: we think of $\Hc$ as constructing 3D space from a collection of concentric spherical shells --- one for each quantised radius. We call $\Hc$ \emph{fuzzy space} or \emph{configuration space}, and write its general element as a ket, $\ket{\psi}$, consistent with the notation above.

\subsection{Quantum state space}\label{sec:Hq}

Our \emph{quantum state (Hilbert) space}, $\Hq$, is defined as the operator algebra acting on $\Hc$ generated by the coordinates. We write its elements $\qket{\psi}$ to distinguish them from those of $\Hc$. The inner product on $\Hq$ is chosen to be the \emph{weighted} Hilbert-Schmidt inner product~\footnote{
	This inner product is chosen so that the resulting norm of the projection operator, $\hat{P}_N$, onto the subspace $\mathcal{F}_0\oplus\cdots\oplus\mathcal{F}_N\subseteq\Hc$ tends to the volume, $\frac{4}{3}\pi r^3$, of the sphere of radius $r=\lambda(N+1)$, as $N\to\infty$~\cite{galikova2013coulomb}.
},
\begin{equation}\label{eq:qip}
	\qip{\psi}{\phi}\defeq 4\pi\lambda^2 \Trc\left(\psi^\dagger\hat{r}\phi\right)
	= 4\pi\lambda^3\Trc\left(\psi^\dagger(\hat{n}+1)\phi\right),
\end{equation}
where the trace $\Trc$ is performed over configuration space $\Hc$.

We can characterise $\Hq$ as the space of Hilbert-Schmidt operators on $\Hc$ which commute with the Casimir operator~\cite{scholtz2018classical}, and express it in the form
\begin{equation}\label{eq:Hq}
	\begin{split}
		\Hq = &\left\{\psi\equiv\sum_{m_i,n_i}C^{m_1,m_2}_{n_1,n_2}
		(a_1^\dagger)^{m_1} (a_2^\dagger)^{m_2} a_1^{n_1} a_2^{n_2} \right.\\
		&\qquad \left|\ \vphantom{\sum_{m_i,n_i}}
		\norm{\psi}<\infty, \quad m_1 + m_2 = n_1 + n_2\right\},
	\end{split}
\end{equation}
where $\norm{\psi}\defeq\qbraket{\psi}$ uses the inner product of \eqref{eq:qip}. The condition $m_1+m_2=n_1+n_2$ ensures commutation with $\hat{r}^2$.

We also define the larger space, $\mathscr{B}_2(\Hc)\supset\Hq$, containing \emph{all} Hilbert-Schmidt operators on $\Hc$ (with identical inner product), which is also a Hilbert space~\cite{conway2019course}. Indeed, we get an isometric isomorphism (see section 2.6 of~\cite{kadison1986fundamentals})
\begin{equation}\label{eq:B2Hc}
	\begin{split}
		\mathscr{B}_2(\Hc) &\cong \Hc\otimes\Hc^* \\
		&= \spn\left\{\ketbra{m_1,m_2}{n_1,n_2}
		\ \middle|\ n_i,m_i\in\mathbb{N}\right\},
	\end{split}
\end{equation}
up to completion with respect to the norm.

The restriction of having to commute with $\hat{r}^2$ forces operators in $\Hq$ to restrict to each subspace $\mathcal{F}_n$ in \eqref{eq:Hc-irreps}. Explicitly this means
\begin{equation}\label{eq:Hq-as-direct-sum}
	\Hq = \bigoplus_{n\in\mathbb{N}} \mathscr{B}_2(\mathcal{F}_n)
	\subset \mathscr{B}_2\!\left(\bigoplus_{n\in\mathbb{N}}\mathcal{F}_n\right)
	= \mathscr{B}_2(\Hc),
\end{equation}
so that, in the notation of \eqref{eq:B2Hc},
\begin{equation}\label{eq:Hq-alt}
	\Hq = \spn\left\{\ketbra{m_1,m_2}{n_1,n_2}
	\ \middle|\ n_1 + n_2 = m_1 + m_2\right\},
\end{equation}
where we recall from \eqref{eq:Hc-irreps} the constraint defining each $\mathcal{F}_n$. Of course, each ket $\ket{n_1,n_2}\in\Hc$ is also a simultaneous eigenket of $\hat{r}^2=\hat{x}_\mu\hat{x}_\mu$ and $\hat{x}_3$:
\begin{equation}\label{eq:jm}
	\begin{alignedat}{2}
		\hat{r}^2\ket{n_1,n_2} &= 4\lambda^2\, j(j+1)\ket{n_1,n_2},
		\quad&&\text{where } j=\frac{n_1+n_2}{2}\\
		\hat{x}_3\ket{n_1,n_2} &= 2\lambda\, m\ket{n_1,n_2},
		&&\text{where } m=\frac{n_1-n_2}{2}.
	\end{alignedat}
\end{equation}
allowing us to alternately label using quantum numbers $j$ and $m$, which, wherever they appear, (implicitly) range over $j\in\mathbb{N}/2$ and $m\in\mathbb{Z}/2\cap [-j,j]$. In this notation, \eqref{eq:Hq-alt} is yet simpler:
\begin{equation}\label{eq:Hq-alt2}
	\Hq = \spn\left\{\ketbra{j=n/2,m}{j=n/2,m'}
	\ \middle|\ n\in\mathbb{N}\right\}.
\end{equation}

\subsection{Observables}

As usual, observables are Hermitian operators on $\Hq$. We begin by seeking operators $\hat{X}_i$ and $\hat{L}_j$ for position and angular momentum respectively (we use capital letters to distinguish observables from operators on $\Hc$).

Position operators $\hat{X}_i$ act as usual by left multiplication,
\begin{equation}\label{eq:X}
	\hat{X}_i\qket{\psi} \defeq\qket{\hat{x}_i\psi}.
\end{equation}
given that they should share the commutation relations of the coordinates $x_i$. We also lift the radius operator $\hat{r}$ to an observable by the same multiplicative action
\begin{equation}\label{eq:R-op}
	\hat{R}\qket{\psi}\defeq\qket{\hat{r}\psi}.
\end{equation}
Angular momentum operators $\hat{L}_i$ have the adjoint action,
\begin{equation}\label{eq:L}
	\hat{L}_i\qket{\psi}\defeq\qket{\frac{\hbar}{2\lambda}\left[\hat{x}_i,\psi\right]},
\end{equation}
which is easily shown to yield the normal angular momentum commutation relations.

Some other observables are also straightforward generalisations of their commutative counterparts. For instance, the Hamiltonian takes the familiar form
\begin{equation}\label{eq:H}
	\hat{H} = -\frac{\hbar^2}{2m}\hat{\Delta} + V(\hat{R}),
\end{equation}
only now the Laplacian is defined
\begin{equation}\label{eq:laplacian}
	\hat{\Delta}\qket{\psi} \defeq
	-\qket{\frac{1}{\lambda\hat{r}}[a_\alpha^\dagger,[a_\alpha,\psi]]},
\end{equation}
the form of which is motivated in~\cite{galikova2013coulomb}. In particular, each $\hat{L}_i$ commutes with $\hat\Delta$, and hence with $\hat{H}$, ensuring angular momentum conservation.

Another conserved observable, defined on the entirety of $\mathscr{B}_2(\Hc)$, is
\begin{equation}\label{eq:Gamma}
	\hat{\Gamma}\qket{\psi} \defeq \qket{\left[\hat{n},\psi\right]},
\end{equation}
whose eigenstates, given by
\[
	\hat{\Gamma}\ketbra{m_1,m_2}{n_1,n_2} = (m_1+m_2-n_1-n_2)\ketbra{m_1,m_2}{n_1,n_2},
\]
manifestly span (a dense subset of) $\mathscr{B}_2(\Hc)$, as noted above. Moreover, \eqref{eq:Hq-alt} implies that the physical state space, $\Hq\subset\mathscr{B}_2(\Hc)$ is the zero eigenspace of $\hat{\Gamma}$. The conservation of $\hat{\Gamma}$ reassures us that physical states remain so under time-evolution.

Finally, the linear operator
\begin{equation}\label{eq:Q}
	\hat{Q} \defeq \frac{1}{2\pi}\int_0^{2\pi}e^{i\phi\hat{\Gamma}}\,d\phi,
\end{equation}
is easily seen by the spectral theorem to project from $\mathscr{B}_2(\Hc)$ onto $\ker\hat\Gamma\cong\Hq$. This projection proves useful in section \ref{sec:pos-measurement} for defining position measurement.

\subsection{Position measurement POVM}\label{sec:pos-measurement}

One remaining subtlety in our NCQM framework is the notion of position measurement. Since our position operators $\hat{X}_i$ do not commute, they have no simultaneous eigenbasis, and so the notion of a position eigenstate must be replaced with a \emph{minimal uncertainty} state. It is well known~\cite{klauder1985coherent} that the Glauber coherent states,
\[
	\ket{\bm{z}} \equiv \ket{z_1,z_2}
	= e^{-\frac{1}{2}\bar{z}_\alpha z_\alpha} e^{z_\alpha a_\alpha^\dagger}\ket{0},
	\text{ where }
	\bm{z} = \begin{bmatrix}z_1\\z_2\end{bmatrix}\in\mathbb{C}^2,
\]
suit this purpose. Their pertinent properties are found in~\cite{klauder1985coherent}.

To perform a position measurement at a point,
\(
\bm{D} \equiv (r,\theta,\phi),
\)
we encode the coordinates of $\bm{D}$ in the values
\begin{equation}\label{eq:z-def}
	\begin{split}
		z_1 &= \sqrt{\frac{r}{\lambda}}\, \cos(\frac{\theta}{2})
		e^{-i\frac{\phi}{2}}e^{i\gamma}\\
		z_2 &= \sqrt{\frac{r}{\lambda}}\, \sin(\frac{\theta}{2})
		e^{i\frac{\phi}{2}}e^{i\gamma}.
	\end{split}
\end{equation}
of a coherent state $\ket{\bm{z}}$. Hereafter we adopt the notation
\(
R \defeq \frac{r}{\lambda} = \bar{z}_\alpha z_\alpha,
\)
for the dimensionless radius. The encoding is such that the expectation values,
\(
\expval{\hat{x}_i}{\bm{z}} = \lambda\, \bm{z}^\dagger\sigma^i\bm{z},
\)
reproduce the coordinates of $\bm{D}$:
\begin{equation*}
	\begin{split}
		x_1 &\equiv \expval{\hat{x}_1}{\bm{z}} = r\sin\theta\cos\phi\\
		x_2 &\equiv \expval{\hat{x}_2}{\bm{z}} = r\sin\theta\sin\phi\\
		x_3 &\equiv \expval{\hat{x}_3}{\bm{z}} = r\cos\theta.
	\end{split}
\end{equation*}
Notably, the global phase $\gamma$ drops out of each of the above (as it must), and so constitutes an additional degree of freedom when choosing the $z_i$.

This notion of position exists on the level of configuration space, $\Hc$; to lift it to  the quantum Hilbert space, $\Hq$, we introduce corresponding states $\qket{z_1,z_2,n_1,n_2}_\text{ph}$, defined
\begin{equation}\label{eq:POVM-quantum-states}
	\qket{z_1,z_2,n_1,n_2}_\text{ph} \defeq \hat{Q} \frac{1}{\sqrt{4\pi\lambda^2\hat{r}}}
	\ketbra{\bm{z}}{n_1,n_2}
\end{equation}
where the inverse square-root is included for normalisation and the projection $\hat{Q}$ must be applied to obtain physical states.

We now define position measurement using a Positive-Operator-Valued Measure (POVM). The relevant operators are
\begin{equation}\label{eq:POVM}
	\hat\Pi_{\bm{z}} \defeq \sum_{n_1,n_2} \qket{z_1,z_2,n_1,n_2}_\text{ph}
	\prescript{}{\text{ph}}{\qbra{z_1,z_2,n_1,n_2}},
\end{equation}
which clearly satisfy the requirements of a POVM: they are Hermitian, positive semi-definite (but not orthogonal), and satisfy a completeness relation (readily deduced from the completeness of coherent states),
\begin{equation}\label{eq:POVM-completeness}
	\int\frac{d^4z}{\pi^2}\, \hat\Pi_{\bm{z}} = \hat{Q} = \idq,
\end{equation}
where the measure $d^4z$, sometimes written $d\bar{z}_1dz_1d\bar{z}_2dz_2$, is shorthand for $d\Re(z_1)d\Im(z_1)d\Re(z_2)d\Im(z_2)$.

Finally, the probability density function (PDF) associated with measuring a particle having initial density matrix $\rho$ at the point $\bm{D}$ is given by the usual Born rule,
\begin{equation}\label{eq:pos-Born-rule}
	P(\bm{D}) = \Tr_\mathrm{q}\left(\hat\Pi_{\bm{z}}\rho\right).
\end{equation}
The subscript on $\Tr_\mathrm{q}$ emphasises that the trace is taken over $\Hq$.

Of particular interest to us, the special case of an initially-pure state $\rho=\qketbra{\psi}$, reduces \eqref{eq:pos-Born-rule} to
\begin{equation}\label{eq:pos-Born-rule-pure-state}
	P(\bm{D}) = 4\pi\lambda^2 \ev{\psi\hat{r}\psi^\dagger}{\bm{z}}.
\end{equation}

As expected, $P(\bm{D})$ is independent of the global phase $\gamma$ in \eqref{eq:z-def}, and furthermore the normalisation of $\qket{\psi}$ implies the normalisation of the probability distribution:
\begin{align*}
	1 & = \qip{\psi}{\psi} = 4\pi\lambda^2\Trc\left(\psi^\dagger\hat{r}\psi\right)  \\
	  & = 4\pi\lambda^2\int\frac{d^4z}{\pi^2} \ev{\psi\hat{r}\psi^\dagger}{\bm{z}}.
\end{align*}

Of course, the normalised state $\psi$ has dimension length$^{-3/2}$, so $P(\bm{D})$ is dimensionless. We may therefore wonder how it relates to a \emph{spatial} probability density with dimension length$^{-3}$. For this we rewrite the integration measure in terms of explicit coordinates $(r,\theta,\phi,\gamma)$ (computing the relevant Jacobian from \eqref{eq:z-def}),
\[
	\frac{d^4z}{\pi^2} = \frac{r\sin\theta}{8\pi^2\lambda^2}\,drd\theta d\phi d\gamma,
\]
whereupon
\[
	\int\frac{d^4z}{\pi^2} P(\bm{D}) = \int r^2\sin\theta\,drd\theta d\phi\,
	\frac{\ev{\psi\hat{r}\psi^\dagger}{\bm{z}}}{r},
\]
giving ultimately the interpretation of
\begin{equation}\label{eq:spatial-density}
	\frac{1}{4\pi r\lambda^2} P(\bm{D})
	\equiv \frac{\ev{\psi\hat{r}\psi^\dagger}{\bm{z}}}{r}
\end{equation}
as the spatial density.

\subsection{Coordinate representation}\label{sec:wavefunctions}

We have yet to formulate the analogues of wavefunctions --- that is, coordinate representations of states. We saw in section~\ref{sec:pos-measurement} how coherent states encode positions. If we could lift this position encoding from coherent states $\ket{\bm{z}}\in\Hc$ to corresponding states $\qket{\bm{z}}\in\Hq$, we could define the coordinate representation of a state $\qket{\psi}\in\Hq$ by
\(
\psi(\bm{z}) \defeq \qip{\bm{z}}{\psi},
\)
resembling the approach of commutative quantum mechanics. The states $\qket{\bm{z}}$ should also be complete to readily obtain square-integrability of $\psi(\bm{z})$.

In section~\ref{sec:pos-measurement}, we already encountered one possible way of ``lifting'' coherent states to $\Hq$, namely the states $\qket{z_1,z_2,n_1,n_2}_\text{ph}$ of \eqref{eq:POVM-quantum-states}. Indeed, these states possess some desiderata we now seek to fulfil --- they lift the coordinate encoding to $\Hq$, and satisfy completeness relation \eqref{eq:POVM-completeness} --- but they carry an undesired additional dependence on $n_1$ and $n_2$. Eliminating these labels will force us to adopt a non-trivial product between coordinate-represented states (which must of course be non-commutative).

To this end, first rewrite the trace in \eqref{eq:POVM} instead as an integral running over coherent states, obtaining
\begin{equation}\label{eq:POVM-as-integral}
	\hat\Pi_{\bm{z}} = \int\frac{d^4w}{\pi^2}
	\qket{\bm{z},\bm{w}}_\text{ph}
	\prescript{}{\text{ph}}{\qbra{\bm{z},\bm{w}}},
\end{equation}
for states
\[
	\qket{\bm{z},\bm{w}}_\text{ph} \equiv \qket{z_1,z_2,w_1,w_2}_\text{ph}
	\defeq \hat{Q}\frac{1}{\sqrt{4\pi\lambda^2\hat{r}}}	\ketbra{\bm{z}}{w_1,w_2}.
\]
Next, we can shift the integral in \eqref{eq:POVM-as-integral} with the substitution $\bm{v} \defeq \bm{w} - \bm{z}$, and rewrite the result in terms of translation operators of the form $e^{v_\alpha\partial_{z_\alpha}}$. This simplifies to
\(
\hat\Pi_{\bm{z}} = \qket{\bm{z}} \bar\star \qbra{\bm{z}},
\)
where the states $\qket{\bm{z}} \equiv \qket{\bm{z},\bm{z}}_\text{ph}$ are exactly those we seek, and where we have introduced the ``star-product'',
\begin{align*}
	\bar\star \defeq \int\frac{d^4v}{\pi^2}\,
	e^{-\bar{v}_\alpha v_\alpha}
	e^{\bar{v}_\alpha \overleftarrow{\partial}_{\!\bar{z}_\alpha}}
	e^{v_\alpha \overrightarrow{\partial}_{\!z_\alpha}}
	= \exp[\overleftarrow{\partial}_{\!\bar{z}_\alpha}
	\overrightarrow{\partial}_{\!z_\alpha}].
\end{align*}
In fact, this product is exactly the conjugate (treating differential operators as Wirtinger derivatives~\cite{kaup2011holomorphic}) of the well-known Voros product (2.16 in~\cite{alexanian2001generalized}),
\(
\star \defeq \exp[\overleftarrow{\partial}_{\!z_\alpha}
\overrightarrow{\partial}_{\!\bar{z}_\alpha}],
\)
as suggested by the overbar. In this notation, the completeness relation of \eqref{eq:POVM-completeness} becomes
\begin{equation}\label{eq:star-completeness}
	\int\frac{d^4z}{\pi^2} \qket{\bm{z}} \bar\star \qbra{\bm{z}} = \idq.
\end{equation}

In summary, we define the coordinate representation of a state $\qket{\psi}$ by
\begin{equation}\label{eq:wavefunction}
	\psi(\bm{z}) \defeq \qip{\bm{z}}{\psi}
	= \expval{\sqrt{4\pi\lambda^2\hat{r}}\, \psi}{\bm{z}}.
\end{equation}
Though seemingly a function of two complex variables, by parameterising $z_1$ and $z_2$ as in \eqref{eq:z-def} we find that $\psi(\bm{z})$ is independent of $\gamma$, making it a function on
\(
\mathbb{C}^2/\mathrm{U}(1) \cong \mathbb{R}^3,
\)
as expected. The expected square-integrability of the coordinate representation is automatically implied by the finiteness of the operator norm, $\norm{\psi}$; we simply invoke \eqref{eq:star-completeness} to insert the identity in the operator inner product:
\begin{equation*}
	\qip{\psi}{\psi}
	= \int\frac{d^4z}{\pi^2} \qip{\psi}{\bm{z}} \bar\star \qip{\bm{z}}{\psi}
	= \int\frac{d^4z}{\pi^2}\,
	\bar\psi(\bm{z})\, \bar\star\, \psi(\bm{z}) < \infty.
\end{equation*}

Whenever we care about the radial dependence of the coordinate representation, we must account for the factor of $\sqrt{\hat{r}}$ in \eqref{eq:wavefunction} coming from our weighted inner product. In such cases, we can normalise $\psi$ with a factor of $\frac{1}{\sqrt{4\pi\lambda^2\hat{r}}}$ before computing the coordinate representation, which then reduces to the simple expectation value,
\(
\psi(\bm{z}) = \ev{\psi}{\bm{z}},
\)
called the \emph{symbol} of $\psi$.

\subsection{Position measurement as weak measurement}

Pittaway \emph{et al.}~\cite{pittaway2021quantum} give a detailed account of how it is possible to interpret our POVM framework for position measurement as a form of \emph{weak measurement}. It is worth briefly mentioning here (deferring to the reference for details), mainly because it preempts the quantum-to-classical transition presented in section~\ref{ch:pinhole}, but also because it highlights how position measurement is distinguished from other kinds of measurement within our formalism.

Recall \eqref{eq:Hq-as-direct-sum}, which decomposed our state space as
\[
	\Hq = \bigoplus_{n\in\mathbb{N}} \mathcal{F}_n \otimes \mathcal{F}_n^*
	\subset \Hc \otimes \Hc^*.
\]
Now position observables $\hat{X}_i$ act via left-multiplication (as per \eqref{eq:X}), and thereby only act on one sector, $\Hc$, of $\Hq$. Consequently, position measurements are \emph{local} measurements, only capable of providing information on this sector. The other sector, $\Hc^*$, acts as an \emph{environment} (in the sense of decoherence; see~\cite{schlosshauer2019quantum}, for instance), providing additional degrees of freedom that remain unprobed by position measurements.

That is to say, upon performing a local measurement such as a position measurement, the environmental degrees of freedom can be traced out (by partial trace over the unobserved sector, $\Hc^*$). The result is a post-measurement reduced density matrix, which is generally an improper mixed state. This entirely mirrors the mechanism of decoherence.

The parallel is noteworthy because decoherence is well-understood to generally give rise to suppression of interference terms and thereby to emergent classical behaviour~\cite{schlosshauer2019quantum}.

\section{Free particle solutions}\label{ch:free-particle}

Analogous to the commutative case, the time independent free particle Schr\"odinger equation reads
\begin{equation}\label{eq:free-schrodinger}
    \hat{H}\qket{\psi} = -\frac{\hbar^2}{2m}\hat{\Delta}\qket{\psi} = E\qket{\psi}.
\end{equation}
As in the commutative case, we can obtain both plane wave and radial (spherical wave) solutions. Both forms are needed in our investigation, so we consider each in turn. The discussion of this section adds detail to, but otherwise mimics, that in~\cite{kriel2017scattering}.

\subsection{Non-commutative plane waves}\label{sec:plane-waves}

A natural candidate for the form of a wave solution is
\begin{equation}\label{eq:plane-waves}
    \qket{\bm{p}} \equiv \qket{\bm{k}}
    \defeq \exp[\frac{i}{\hbar}\,\bm{p}\cdot\bhat{x}]
    = e^{i\bm{k}\cdot\bhat{x}},
\end{equation}
up to appropriate normalisation (of which we defer discussion to section~\ref{sec:plane-wave-normalisation}), and where $\bm{k}=\bm{p}/\hbar$ is the de Broglie wave-vector.

Since the $\hat{x}_i$ represent $\su2$ elements (and since $\SU2$ is simply-connected~\cite{hall2003lie}), such plane waves are representations of $\SU2$ group elements. Moreover, the exponential map $\exp\colon\su2\to\SU2$ is surjective, (because $\SU2$ is connected and compact --- see, for instance, corollary 11.10 in~\cite{hall2003lie}), so that \emph{every} (represented) $\SU2$ element assumes the form of \eqref{eq:plane-waves}. A consequence is that composing plane waves produces another plane wave:
\[
    e^{i\bm{k}_1\cdot\bhat{x}}e^{i\bm{k}_2\cdot\bhat{x}}
    = e^{i\bm{k}_3\cdot\bhat{x}}.
\]
We can compute the momentum $\bm{k}_3$ of this new plane wave using the well-known Baker–Campbell–Hausdorff (BCH) expansion for $\log(e^Xe^Y)$, which has a simple closed form in the case of $\SU2$ (attributed to Rodrigues~\cite{rodrigues1840lois}). For this, first rewrite the plane waves in terms of dimensionless quantities $\kappa_i$ as
\begin{equation}\label{eq:plane-wave-dimensionless}
    e^{i\bm{k}_i\cdot\bhat{x}}
    \equiv e^{i\kappa_i\,\bhat{k}_i\cdot\frac{\bhat{x}}{\lambda}},
    \quad\text{where}\quad\kappa_i\defeq\lambda\norm{\bm{k}_i},
\end{equation}
for unit vectors $\bhat{k}_i$. Then solve for $\kappa_3$ and $\bhat{k}_3$ using
\begin{equation}\label{eq:SU2-BCH}
    \begin{split}
        \cos\kappa_3 &= \cos\kappa_1\cos\kappa_2
        - \bhat{k}_1\cdot\bhat{k}_2\sin\kappa_1\sin\kappa_2,\\
        \bhat{k}_3 &= \frac{1}{\sin\kappa_3} \left(\bhat{k}_1\sin\kappa_1\cos\kappa_2
        + \bhat{k}_2\sin\kappa_2\cos\kappa_1\right.\\&\qquad\qquad\qquad
        \left. - \bhat{k}_1\times\bhat{k}_2\sin\kappa_1\sin\kappa_2\right).
    \end{split}
\end{equation}
Since the Pauli matrices share commutation relations with the operators $\hat{x}_i/\lambda$, an analogous composition rule holds for (unrepresented) $\SU2$ group elements written as exponentials in this way.

Plane waves transform simply under the action of a rotation operator. Let $R\equiv R_\phi(\bhat{u})\in\mathrm{SO}(3)$ be the rotation matrix with angle $\phi$ and axis $\bhat{u}$, and let
\[
    \Pi(R) \defeq \exp[-\frac{i}{\hbar}\,\phi\bhat{u}\cdot\bhat{L}]
\]
be its representation on $\Hq$. For convenience define $\hat{J}_i \equiv \frac{1}{2\lambda}\hat{x}_i$, and note that
\(
[\hat{x}_i,\hat{J}_j] = i\epsilon_{ijk}\hat{x}_k = [\hat{J}_i,\hat{J}_j],
\)
implying that $\bhat{x}$ is a \emph{vector operator} with respect to the $\hat{J}_i$, from which it immediately follows by the Baker-Hausdorff lemma (see section 5.1.2 of~\cite{abers2003quantum}, for instance) that
\[
    e^{-i\phi\bhat{u}\cdot\bhat{J}}\,\hat{x}_i\,
    e^{i\phi\bhat{u}\cdot\bhat{J}}
    = \left[R_{-\phi}(\bhat{u})\right]_{ij}\hat{x}_j
    = \left[R^T\bhat{x}\right]_{i}.
\]
Combining the above with definition \eqref{eq:L},
\(
\hat{L}_i \defeq \ad_{\hbar\hat{x}_i / 2\lambda},
\)
it is then easily seen that
\begin{equation}\label{eq:rotation-of-plane-wave}
    \begin{split}
        \Pi(R)\qket{\bm{k}} &= \exp[\ad_{-i\phi\bhat{u}\cdot\bhat{J}}]
        e^{i\bm{k}\cdot\bhat{x}}\\
        &= \Ad_{\exp[-i\phi\bhat{u}\cdot\bhat{J}]}
        \left(e^{i\bm{k}\cdot\bhat{x}}\right)\\
        &= \qket{R\bm{k}},
    \end{split}
\end{equation}
where we have used $e^{\ad_X}=\Ad_{e^X}$ (see 3.34 in Hall~\cite{hall2003lie}, for instance).

For the purposes of computing the plane wave energy, we use the above rotation law together with the rotational invariance of the Schr\"odinger equation, \eqref{eq:free-schrodinger}, to focus on the case $\bm{k}=k\bhat{z}$,
\begin{equation}\label{eq:plane-wave-z}
    \qket{k\bhat{z}} = e^{ik\hat{x}_3}
    = e^{ik\lambda\left(a_1^\dagger a_1 - a_2^\dagger a_2\right)}
    \equiv e^{ik\lambda\left(\hat{n}_1 - \hat{n}_2\right)}.
\end{equation}
Now the energy is readily computed directly from the commutation relations
\(
f(\hat{n}_j)a_j = a_j f(\hat{n}_j-1)
\)
(specifically with
\(
f(\hat{n}_1, \hat{n}_2) = e^{ik\lambda\left(\hat{n}_1 - \hat{n}_2\right)}
\)%
) and $a_ja_j^\dagger = \hat{n}_j + 1$, as such:
\begin{equation}\label{eq:plane-wave-energy}
    \begin{split}
        \hat{H}\qket{k\bhat{z}} &= \frac{\hbar^2}{2\lambda m\hat{r}}
        \left[a_\alpha^\dagger, \left[a_\alpha,
                e^{ik\lambda\left(\hat{n}_1 - \hat{n}_2\right)}\right]\right]\\
        &= \frac{2\hbar^2}{m\lambda^2}
        \sin^2\left(\frac{k\lambda}{2}\right) \qket{k\bhat{z}}.
    \end{split}
\end{equation}
The non-commutativity has clearly affected the usual dispersion relation, in particular introducing an energy upper bound
\(
E_{\text{max}} = \frac{2\hbar^2}{m\lambda^2}.
\)
This is consistent with the restriction $k\in[0,\pi/\lambda)$ on $k$ required for the states in \eqref{eq:plane-wave-z} to be linearly independent.

\subsection{Plane wave normalisation}\label{sec:plane-wave-normalisation}

We take some time to motivate our choice of plane wave normalisation. Firstly, note that, being eigenstates of a Hermitian operator (as per \eqref{eq:plane-wave-energy}), plane waves with different energies are necessarily orthogonal. Equivalently, plane waves $\qket{\bm{k}_1}$ and $\qket{\bm{k}_2}$ with non-trivial overlap must satisfy
\begin{align*}
         & \sin^2\left(\frac{k_1\lambda}{2}\right)
    = \sin^2\left(\frac{k_2\lambda}{2}\right)      \\
    \iff & k_1 = \pm k_2 + m \frac{2\pi}{\lambda},
    \quad\text{for some}\quad m\in\mathbb{Z},
\end{align*}
which is at least consistent with the requirement that wavenumbers be restricted to the interval $[0,\pi/\lambda)$ for linear independence.

To aid in selecting an appropriate normalisation condition, we explicitly compute the overlap of two plane waves. By \eqref{eq:qip}, together with \eqref{eq:plane-waves}, we have
\begin{equation*}
    \qip{\bm{k}_1}{\bm{k}_2} = 4\pi\lambda^3 \Trc\left(
    e^{-i\bm{k}_1\cdot\bhat{x}} (\hat{n}+1)\, e^{i\bm{k}_2\cdot\bhat{x}}\right).
\end{equation*}
Now plane waves commute with $\hat{n}$ (as each coordinate does), and we can compose the plane waves using \eqref{eq:SU2-BCH} to obtain
\begin{equation*}
    \qip{\bm{k}_1}{\bm{k}_2} = 4\pi\lambda^3 \sum_{j,m} (2j + 1)
    \ev{e^{i\bm{k}_3\cdot\bhat{x}}}{j,m},
\end{equation*}
for some new wave-vector $\bm{k}_3$. Here, $\ket{j,m}$ is a simultaneous eigenket of
$\hat{r}^2$ and $\hat{x}_3$, as in \eqref{eq:jm}. In this expression we recognise the \emph{character} of the $j$th $\SU2$ irrep, usually written
\[
    \chi^j(\bm{k}_3) \equiv \chi^j(\alpha,\beta,\gamma)
    \defeq \sum_m D^j_{mm}(\alpha,\beta,\gamma),
\]
in terms of the Wigner-$D$ matrix element,
\[
    D^j_{m',m}(\alpha,\beta,\gamma)	\defeq \mel{j,m'}{e^{i\alpha\frac{\hat{x}_3}{\lambda}}
        e^{i\beta\frac{\hat{x}_2}{\lambda}} e^{i\gamma\frac{\hat{x}_3}{\lambda}}}{j,m}.
\]
This usual parameterisation uses Euler angles $(\alpha,\beta,\gamma)$, relying on the decomposition
\(
e^{i\bm{k}_3\cdot\bhat{x}}
\equiv e^{i\alpha\frac{\hat{x}_3}{\lambda}}
e^{i\beta\frac{\hat{x}_2}{\lambda}} e^{i\gamma\frac{\hat{x}_3}{\lambda}},
\)
of our plane wave. Importantly, characters are invariant on conjugacy classes, due to the cyclic property of the trace. In particular, $\chi^j(\bm{k}_3)$ depends only on the magnitude $k_3$, as we may conjugate with another plane wave to arbitrarily rotate the wave-vector direction (as in \eqref{eq:rotation-of-plane-wave}) without affecting the character. Thus, we will opt to rotate $\bm{k}_3$ to lie entirely along the $\hat{x}_2$-axis, whence we can take $\alpha = \gamma = 0$ and $\beta = \kappa_3 \defeq \lambda k_3$. Finally, we can invoke the completeness relation for characters (equation 3.95 in~\cite{schwinger1952angular}) to write
\begin{equation}\label{eq:plane-wave-overlap}
    \qip{\bm{k}_1}{\bm{k}_2} = 4\pi\lambda^3\ \delta(\kappa_3),
\end{equation}
since $\chi^j(0)$ is the dimension of the $j$th $\SU2$ irrep, $2j+1$.

We confirm that \eqref{eq:plane-wave-overlap} is sensible by determining the conditions leading to non-zero overlap. The overlap is clearly non-zero exactly when $\kappa_3=0$, whereupon the composition
\(
\left(e^{i\bm{k}_1\cdot\bhat{x}}\right)^\dagger e^{i\bm{k}_2\cdot\bhat{x}}
\)
of our original plane waves gives the identity operator
\(
\idc \equiv e^{0\,\bhat{k}_3\cdot\frac{\bhat{x}}{\lambda}}.
\)
Because our $\SU2$ representation is \emph{faithful}, this happens precisely when the corresponding (unrepresented) $\SU2$ group elements compose to the identity matrix,
\[
    \left(e^{i\lambda\bm{k}_1\cdot\bhat\sigma}\right)^\dagger
    e^{i\lambda\bm{k}_2\cdot\bhat\sigma}
    = I,
\]
\textit{i.e.}~when the unrepresented group elements are equal. This translates into a slightly non-trivial condition on $\bm{k}_1$ and $\bm{k}_2$, since the exponential map $\exp\colon\su2\to\SU2$ is not injective~\cite{hall2003lie}. To make this more precise, we first expand both exponentials (using \eqref{eq:expanded-su2-element} from appendix~\ref{app:su2-transformations}),
\[
    \cos(\lambda k_1) I + i\sin(\lambda k_1)\bhat{k}_1\cdot\bhat\sigma
    = \cos(\lambda k_2) I + i\sin(\lambda k_2)\bhat{k}_2\cdot\bhat\sigma,
\]
then note that the matrices $\left\{I, \hat\sigma_1, \hat\sigma_2, \hat\sigma_3\right\}$ form a basis for $\mathcal{M}_{2,2}(\mathbb{C})$, and so are linearly independent, whereby we can equate coefficients,
\begin{align*}
    \cos(\lambda k_1)           & = \cos(\lambda k_2),           \\
    \sin(\lambda k_1)\bhat{k}_1 & = \sin(\lambda k_2)\bhat{k}_2.
\end{align*}
It follows that $\qip{\bm{k}_1}{\bm{k}_2}$ is non-zero exactly when $k_1 \pm k_2\in\frac{2\pi}{\lambda}\mathbb{Z},$ and either
\begin{enumerate}
    \item $k_1\in\frac{\pi}{\lambda}\mathbb{Z},$ or
    \item $\bhat{k}_1 =\mp\bhat{k}_2$.
\end{enumerate}
If we restrict $k_1,k_2\in[0,\pi/\lambda)$ then these cases collapse into the much simpler (expected) single condition that $\bm{k}_1 = \bm{k}_2$.

The conclusion is that we may simply choose
\begin{equation}\label{eq:normalisation-constant}
    N = \frac{1}{\sqrt{4\pi\lambda^3}},
\end{equation}
to normalise our plane waves, so that the normalisation condition reads
\begin{equation}\label{eq:normalisation-condition}
    \qip{\bm{k}_1}{\bm{k}_2} = \delta(\kappa_3),
\end{equation}
where $\cos\kappa_3$ is given by \eqref{eq:SU2-BCH}. This normalisation condition, while unconventional, is both convenient and sensible, as shown. It also gives $\qket{\bm{k}}$ the correct dimensions of length$^{-3/2}$, as per the discussion of dimensions at the end of section~\ref{sec:pos-measurement}.

\subsection{Commutative spherical waves}
\label{sec:commutative-spherical-waves}

Spherical waves play an important role in the pinhole interference setup which we study in section~\ref{ch:pinhole}. As such, before we treat the non-commutative free particle spherical wave solutions, it bears revising the standard commutative spherical waves for later comparison. These are derived in many standard introductory texts; see for instance sections 3.3 and 3.4 of Abers~\cite{abers2003quantum}. They have the general form
\[
    \psi_{klm}(r,\theta,\phi) = R_{kl}(r) Y_l^m(\theta, \phi),
\]
labelled by momentum, $k$, and angular momentum quantum numbers $l\in\mathbb{N}$ and $m\in[-l,l]\cap\mathbb{N}$. The wavefunction normalisation implies separate normalisation conditions on the angular and radial components, expressed respectively as
\begin{equation}\label{eq:radial-schrodinger-normalisation}
    \int \abs{Y_l^m(\theta, \phi)}^2\, d\Omega = 1,\quad\text{and}\quad
    \int_0^\infty \abs{R_{kl}(r)}^2\, r^2dr = 1.
\end{equation}
For such solutions, the Schr\"odinger equation separates into angular- and radial equations. The former admits the standard spherical harmonic solutions, which we omit. The latter admits solutions of the form
\[
    R_{kl}(r) = A g_{J,l}(kr) + B g_{Y,l}(kr),
\]
where $g_{J,l}$ and $g_{Y,l}$ denote respectively the order-$l$ spherical Bessel- and Neumann functions, and where the coefficients $A$ and $B$ are given by boundary conditions. The Neumann functions are only valid away from the origin, but this regime is relevant for our calculation. Indeed, we will use the large-$r$ form of the radial solutions (see equation 3.132 of~\cite{abers2003quantum}, for instance),
\begin{equation}\label{eq:asymptotic-spherical-waves}
    R_{kl}(r) = N\, \frac{1}{r} e^{\pm ikr},
\end{equation}
where $N$ is a normalisation constant. This asymptotic form follows from the large-$r$ limiting behaviour of the spherical Hankel functions of the first kind,
\(
g_{H,l} = g_{J,l} + i g_{Y,l};
\)
see section A.2.3 of~\cite{abers2003quantum} for a derivation. This approach will inform how we derive the asymptotic form of the analogous \emph{non-commutative} spherical waves in section~\ref{sec:spherical-waves} below.

Concerning normalisation, there are two important observations to make. Firstly, it is clear from \eqref{eq:radial-schrodinger-normalisation} that the radial part of any wavefunction has dimensions of length$^{-3/2}$; therefore, $N$ should have dimensions of length$^{-1/2}$. Secondly, the integral as given in \eqref{eq:radial-schrodinger-normalisation} diverges if we insert the asymptotic form of $R_{k,l}$,
\[
    \int_0^\infty \abs{N\, \frac{1}{r} e^{\pm ikr}}^2\, r^2dr
    = \abs{N}^2 \int_0^\infty dr \not< \infty.
\]
We address both observations by regularising the divergent integral by bounding our system within a large finite volume $V$, say a sphere, and taking the radial integral only to the boundary of the volume, $\sim V^{1/3}$. Then $N$ is proportional to $V^{-1/6}$, which introduces the missing dimensionality. We will not bother to calculate the exact dimensionless proportionality constant, as it is neither very meaningful (\eqref{eq:asymptotic-spherical-waves} is merely an approximation for large $r$) nor necessary for our discussion.

\subsection{Non-commutative spherical waves}
\label{sec:spherical-waves}

The free particle angular momentum eigenstates are shown in~\cite{galikova2013coulomb} to have the form
\begin{equation}\label{eq:L-eigenstates}
    \qket{k,l,m} \defeq \sum_{(m_i, n_i)\in\Lambda}
    \frac{(a_1^\dagger)^{m_1}(a_2^\dagger)^{m_2}}{m_1!\, m_2!} g(\hat{n},k)
    \frac{(a_1)^{n_1}(-a_2)^{n_2}}{n_1!\, n_2!}.
\end{equation}
Here, $k$ is the momentum, $l\in\mathbb{N}$ and $m\in\mathbb{Z}\cap[-l,l]$ are fixed indices, and the summation runs over the set
\begin{equation}\label{eq:Lambda-set}
    \begin{split}
        \Lambda \defeq &\left\{(m_1,m_2,n_1,n_2)\in\mathbb{N}^4 \ \right|\
        m_1 + m_2 = n_1 + n_2 = l, \\
        &\quad\left.\vphantom{\mathbb{N}^4} m_1 - m_2 - n_1 + n_2 = 2m\right\}.
    \end{split}
\end{equation}
Inserting this form into \eqref{eq:free-schrodinger} yields a difference equation for $g$ which admits two linearly independent solutions (derived in section 7 of~\cite{chandra2014spectrum}) which have the forms
    {\small
        \begin{equation}\label{eq:spherical-Bessel}
            \begin{aligned}
                g_{J,l}(n,k) =
                 & \left[\frac{\sqrt{\pi}\, \sin^{l + 1}\kappa}
                    {2^{l + 1}\, \Gamma\left(\frac{3}{2} + l\right)}\right]
                \cos^n\kappa                                    \\
                 & \times {\prescript{}{2}{F}}_{\!\!1}\!
                \left(\frac{1 - n}{2}, -\frac{n}{2},
                \frac{3}{2} + l, -\tan^2\kappa\right),
            \end{aligned}
        \end{equation}
    }
for $n\geq 0$, and
    {\small
        \begin{equation}\label{eq:spherical-Neumann}
            \begin{aligned}
                g_{Y,l}(n,k) =
                 & \left[-\frac{\sqrt{\pi} (-2)^l \cos^{l + 1}\kappa}
                    {\tan^l\kappa \Gamma\left(\frac{1}{2} - l\right)}\right]
                \frac{n!\, \cos^n\kappa}{\Gamma(2 + 2l + n)}          \\
                 & \times\ {\prescript{}{2}{F}}_{\!\!1}\!
                \left(\frac{-1 - l - n}{2}, -l - \frac{n}{2},
                \frac{1}{2} - l, -\tan^2\kappa\right),
            \end{aligned}
        \end{equation}
    }
for $n>0$, respectively; here the functions
\(
{\prescript{}{2}{F}}_{\!\!1}\!
\)
are the hypergeometric functions, and $\kappa \defeq \lambda k$ is a dimensionless quantity. These solutions are the non-commutative analogues of the spherical Bessel- and Neumann functions, respectively. Like its commutative counterpart, the latter solution, $g_{Y,l}$, is only valid away from the origin, captured by the restriction $n > 0$. Away from the origin, we can linearly combine the solutions as
\[
    g_{H,l} \defeq g_{J,l} + i g_{Y,l},
\]
obtaining the non-commutative analogues for the Hankel functions of the first kind.

We are interested in the asymptotic behaviour of the spherical waves. Following the approach in section D of~\cite{kriel2017scattering}, we first consider the asymptotic behaviour of the radial function, $g$. This involves expressing $g_{Y,l}(n,k)$ and $g_{J,l}(n,k)$ in terms of Jacobi polynomials, using identities 15.3.21 and 15.4.6 from~\cite{abramowitz1964handbook}, then expanding these polynomials for large $n$ (corresponding to large radius) using theorem 8.21.8 from~\cite{szeg1939orthogonal}. The result, as given in~\cite{kriel2017scattering}, is that for large $n$,
\begin{align*}
    g_{J,l}(n,k) & \approx \frac{\sin((n - l - 1)\kappa
        - l\pi/2)}{n^{l+1}},
    \quad\text{and}                                      \\
    g_{Y,l}(n,k) & \approx -\frac{\cos((n - l - 1)\kappa
        - l\pi/2)}{n^{l+1}},
\end{align*}
whence the linear combination $g_{H,l}$ behaves asymptotically like an outgoing radial wave:
\begin{equation}\label{eq:asymptotic-Hankel}
    g_{H,l}(n,k) \approx \frac{e^{i(n + l + 1)\kappa}}{(in)^{l + 1}}.
\end{equation}
Compared to its commutative analogue, there is an additional factor of $n^{-l}$ in this expression. However, this is compensated by the remaining radial dependence in (the other factors of) the spherical wave. To see this, consider the (asymptotic) symbol (see section~\ref{sec:wavefunctions}),
\begin{align*}
    \ev{k,l,m}{\bm{z}}
     & = \!\!\sum_{(m_i, n_i)\in\Lambda}\!\!\!\!\!
    \frac{\bar{z}_1^{m_1} \bar{z}_2^{m_2} z_1^{n_1} (-z_2)^{n_2}}
    {m_1!\, m_2!\, n_1!\, n_2!} \ev{g_{H,l}(\hat{n},k)}{\bm{z}}                 \\
     & \approx R^l \ev{\frac{e^{i(\hat{n} + l + 1)\kappa}}{(i\hat{n})^{l + 1}}}
    {\bm{z}} e^{im\phi}                                                         \\
     & \qquad\times \sum_{(m_i, n_i)\in\Lambda}
    \frac{(-1)^{n_2}}{m_1!\, m_2!\, n_1!\, n_2!}                                \\
     & \qquad\qquad\times \cos^{m_1 + n_1}\!\left(\frac{\theta}{2}\right)
    \sin^{m_2 + n_2}\!\left(\frac{\theta}{2}\right),
\end{align*}
of the state $\qket{k,l,m}$. The radial dependence is shared between the first two factors, so we should compute the remaining matrix element. This calculation ends up being somewhat technical, so we delegate the details to appendix~\ref{app:coherent-state-mels-hankel}. There it is shown that, to leading order in the large-$R$ expansion,
\begin{equation}\label{eq:asymptotic-g-symbol}
    \ev{\frac{e^{i(\hat{n} + l + 1)\kappa}}{(i\hat{n})^{l + 1}}}{\bm{z}}
    \sim \frac{1}{(iR)^{l+1}} e^{R(\cos\kappa - 1) + i R\sin\kappa},
\end{equation}
cancelling the factor of $R^l$ from the angular part of the spherical wave symbol, and leaving an overall radial dependence of $1/R$.

\section{Pinhole Interference}\label{ch:pinhole}

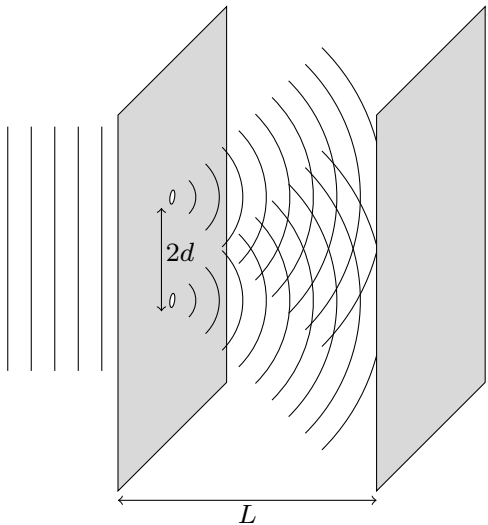
\begin{figure}[!ht]
\centering
\begin{tikzpicture}[scale=1.25,every node/.append style={transform shape}]
	\def\L{2.75} 
	\def\d{0.55} 
	\def\r{0.07} 
	\foreach \x in {0.25,0.5,...,1.5} \draw (\x,-1.3) -- (\x,1.3);
	\draw[fill=black!15] (2,-2,-1.5) -- (2,-2,1.5) -- (2,2,1.5) -- (2,2,-1.5)%
		-- (2,-2,-1.5);
	\begin{scope}[canvas is yz plane at x=2,rotate=-90]
		\draw[fill=white] (0,\d) circle (\r) (0,-\d) circle (\r);
	\end{scope}
	\foreach \r in {0.25,0.5,...,2.25} {
		\draw (2,\d) ++(-45:\r) arc (-45:45:\r);
		\draw (2,-\d) ++(-45:\r) arc (-45:45:\r);
	}
	\draw[fill=black!15] (2+\L,-2,-1.5) -- (2+\L,-2,1.5) -- (2+\L,2,1.5)%
		-- (2+\L,2,-1.5) -- (2+\L,-2,-1.5);
	\draw[<->] (2,\d,0.3) -- (2,-\d,0.3) node[midway,right=-2pt,shift={(0,0.08)}]%
		{\footnotesize $2d$};
	\draw[<->] (2,-2.1,1.5) -- (2+\L,-2.1,1.5) node[midway,below=-2pt]%
		{\footnotesize $L$};
\end{tikzpicture}
\caption{Pinhole interference configuration}
\label{fig:pinhole}
\end{figure}

Consider the pinhole interference configuration in figure \ref{fig:pinhole}. Plane waves incident on a barrier (which is oriented normal to the direction of propagation) pass through a pair of pinholes, resulting in spherical wave fronts that interfere before being detected on a screen. We choose the barrier to lie in the plane $x=0$, with the pinhole apertures at $z=\pm d$.

A point $\bm{D}=(L,y_D,z_D)$ on the screen is then separated from each of the respective pinholes by distances
\begin{equation}
	r_\pm \defeq \dist\left(\pm d\bhat{z},\bm{D}\right)
	\equiv \sqrt{L^2+y_D^2+(z_D\mp d)^2}.
\end{equation}
Due to symmetry, the spherical waves in our setup will have equal energies, and thus (given the usual non-relativistic dispersion relation) equal wavenumber magnitudes, $k$. Of course, spherical waves emanate radially outward in all directions, but we will perform calculations  within the \emph{large-separation approximation}, wherein $L$ is assumed to far exceed both the slit separation, $2d$, and the displacements, $z_D$ and $y_D$, of the measurement point. Under this approximation, we will treat the spherical waves incident on the screen instead as (appropriately scaled) plane waves with momenta normal to their wavefronts. The appropriate wavenumber vectors are then
\begin{equation}
	\bm{k}_{\pm} \equiv k\,\bhat{k}_\pm
	= \frac{k}{r_\pm} \begin{bmatrix}L\\y_D\\z_D\mp d\end{bmatrix},
\end{equation}
directed from each pinhole towards $\bm{D}$. We will see this approximation in effect in the coming sections, both for the commutative and non-commutative calculations.

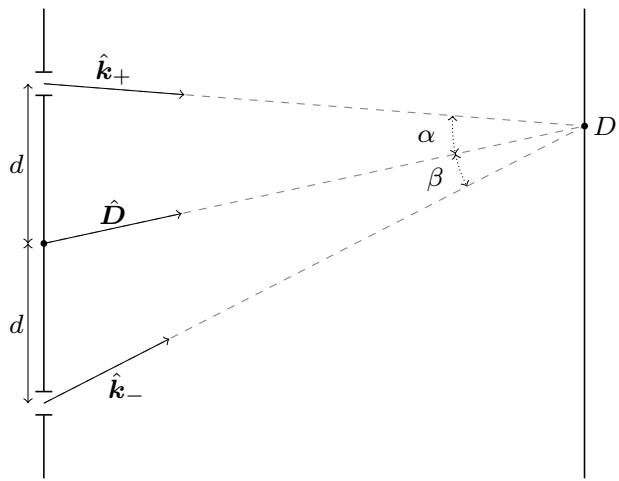
\begin{figure}[!hb]
\centering
\begin{tikzpicture}[scale=1.25,every node/.append style={transform shape}]
	\def\L{5.75} 
	\def\d{1.70} 
	\def\h{2.50} 
	\node (O) at (0,0) {};
	\node (U) at (0,\d) {};
	\node (L) at (0,-\d) {};
	\node[label={[label distance=-4pt]right:{\footnotesize $D$}}] (D) at (\L,1.25) {};
	\draw[-|,semithick] (0,\h) -- (U);
	\draw[|-|,semithick] (U) -- (L);
	\draw[|-,semithick] (L) -- (0,-\h);
	\draw[semithick] (\L,\h) -- (\L,-\h);
	\draw[dashed,gray] (U.center) -- (D.center);
	\draw[dashed,gray] (L.center) -- (D.center);
	\draw[dashed,gray] (O.center) -- (D.center);
	\draw[->] (U.center) -- ($(U.center)!1.5cm!(D.center)$) node[midway,above=-2pt]
		{\footnotesize $\bhat{k}_+$};
	\draw[->] (L.center) -- ($(L.center)!1.5cm!(D.center)$) node[midway,below right=-4pt]
		{\footnotesize $\bhat{k}_-$};
	\draw[->] (O.center) -- ($(O.center)!1.5cm!(D.center)$) node[midway,above=-2pt]
		{\footnotesize $\bhat{D}$};
	\begin{scope}[transform canvas={xshift=-6pt}]
		\draw[<->] (O.center) -- (U.center) node[midway,left=-2pt] {\footnotesize $d$};
		\draw[<->] (O.center) -- (L.center) node[midway,left=-2pt] {\footnotesize $d$};
	\end{scope}
	\pic[draw,densely dotted,<->,"\footnotesize $\alpha$",
		angle eccentricity=1.2,angle radius=40pt] {angle=U--D--O};
	\pic[draw,densely dotted,<->,"\footnotesize $\beta$",
		angle eccentricity=1.2,angle radius=40pt] {angle=O--D--L};
	\draw[fill] (O) circle (0.03);
	\draw[fill] (D) circle (0.03);
\end{tikzpicture}
\caption[Planar slice of pinhole interference configuration]
{Planar slice of pinhole configuration along plane defined by pinholes
and detection point}
\label{fig:pinhole-planar}
\end{figure}

Note also that, since the origin is co-linear with the pinholes, $\bhat{k}_\pm$ and $\bhat{D}$ are all co-planar. This permits us to illustrate the geometry of the setup within an appropriate planar slice; such is given by figure~\ref{fig:pinhole-planar}, whose angles we will reference in the coming calculations.

The main result of this section is derived in subsection~\ref{sec:NCQM-interference-pattern}, where we treat this pinhole configuration in the fuzzy-sphere NCQM formalism. However, in subsection~\ref{sec:commutative-interference-pattern} below, we first treat the setup using ordinary commutative quantum mechanics. This is useful for comparison with the non-commutative result --- indeed, the commutative result should be viewed as a special case of the non-commutative one, and must be entirely recovered within the \emph{commutative limit}, $\lambda\to 0$. We check this limit in section~\ref{sec:commutative-lim}.

\subsection[Commutative calculation]{Commutative interference calculation}
\label{sec:commutative-interference-pattern}

The wavefunction incident on the screen is a superposition of spherical waves emanating from each pinhole. We invoke the large-separation assumption (in particular, that $L \gg 2d$) to use the large-$r$ form of each spherical wave, as per \eqref{eq:asymptotic-spherical-waves}. Then the wavefunction is
\[
	\psi(\bm{D}) = \frac{1}{\sqrt{8\pi}\, V^{1/6}} \left(\frac{1}{r_+}e^{ikr_+}
	+ \frac{1}{r_-}e^{ikr_-}\right),
\]
where $V$ is the system's volume, as per the discussion of section~\ref{sec:commutative-spherical-waves}. We choose the simplest possible angular dependence for our spherical waves, namely the spherical harmonic $Y^0_0(\theta, \phi) = 1/\sqrt{4\pi}$. A more complicated angular dependence would only serve to apply an additional modulation to the base interference pattern --- we therefore simplify the problem by ignoring angular momentum. Crucially, the large-separation assumption further permits us to approximate the spherical waves with (appropriately-attenuated) plane waves,
\begin{equation}\label{eq:state-at-screen-approx}
	\psi(\bm{D}) = \frac{1}{\sqrt{8\pi}\, V^{1/6}}
	\left(\frac{1}{r_+}e^{i\bm{k}_+\cdot\bm{D}}
	+ \frac{1}{r_-}e^{i\bm{k}_-\cdot\bm{D}}\right).
\end{equation}
We should justify this by showing that the plane waves capture the correct radial behaviour within our approximation. To this end, consider the triangle depicted on figure~\ref{fig:pinhole-planar} with vertices $\bm{0}$, $\bm{D}$ and $d\bhat{z}$; its area can be expressed either as $\frac{1}{2}d\sqrt{L^2 + y_D^2}$ or as $\frac{1}{2}r_+r\sin\alpha$, whence
\begin{align*}
	\sin\alpha = \frac{d\sqrt{L^2 + y_D^2}}{rr_+} \approx \frac{d}{L},
\end{align*}
up to first order (in all of $d/L, y_D/L, z_D/L \ll1$). Next, applying the cosine rule,
\(
d^2 = r_+^2 + r^2 - 2rr_+\cos\alpha,
\)
to the same triangle, we get
\begin{equation}\label{eq:cos-rule-quadratic}
	r_+ = r\cos\alpha \pm r\sqrt{\left(\frac{d}{r}\right)^2 - \sin^2\alpha}\,
	\approx r\cos\alpha,
\end{equation}
since $d/r \approx d/L$ also, so the square root vanishes to leading order. By identical reasoning, we also have $r_- \approx r\cos\beta$, and so
\(
\bm{k}_+\cdot\bm{D} = \norm{\bm{k}_+}\norm{\bm{D}}\cos\alpha = rk\cos\alpha \approx k r_+,
\)
and similarly $\bm{k}_-\cdot\bm{D} \approx k r_-$. This justifies the paraxial approximation of \eqref{eq:state-at-screen-approx}.

With this simplified wavefunction, the measurement probability density, $P_\text{comm}(\bm{D}) = \abs{\psi(\bm{D})}^2$, is (up to dimensionless normalisation prefactors)
\begin{equation}\label{eq:P-commutative}
	\begin{split}
		P_\text{comm}(\bm{D}) = \frac{1}{4\pi V^{1/3}\, r_+r_-}
		&\left[\frac{2d^2}{r_+r_-} + \cos(\alpha + \beta) \right. \\
			&\ \ \left.\vphantom{\frac{1}{r_+}}
			+ \cos(rk(\cos\alpha - \cos\beta))\right].
	\end{split}
\end{equation}

\subsection[Non-commutative calculation]{Non-commutative interference calculation}
\label{sec:NCQM-interference-pattern}

Much as in the commutative calculation, we want the state $\qket{\psi}$ arriving at $\bm{D}$ to have the form of a superposition of spherical waves emanating from each pinhole. In terms of symbols,
\[
	\ev{\psi}{\bm{z}} = M \left[\ev{k,0,0}{\bm{z}^+} + \ev{k,0,0}{\bm{z}^-}\right],
\]
where the $\bm{z}^{\pm}\in\mathbb{C}^2$ encode the coordinates, $\bm{D}_\pm \defeq \bm{D} \mp d\bhat{z},$ of $\bm{D}$ (as per \eqref{eq:z-def}) relative to the respective pinholes, and where $M$ is an overall normalisation. We have clearly again chosen $l = 0$ for simplicity, and we will again focus on the large-separation asymptotic behaviour,
\begin{equation}\label{eq:pinhole-spherical-waves}
	\begin{split}
		\ev{\psi}{\bm{z}} \approx \frac{M}{i}
		& \left[\frac{1}{R_+} e^{R_+(\cos\kappa - 1) + iR_+\sin\kappa} \right.        \\
			& \ \ \left. + \frac{1}{R_-} e^{R_-(\cos\kappa - 1) + iR_-\sin\kappa}\right],
	\end{split}
\end{equation}
as per \eqref{eq:asymptotic-g-symbol}, where, of course, we have defined the dimensionless
\(
R_\pm \defeq \frac{r_\pm}{\lambda} = \bar{z}^\pm_\alpha z^\pm_\alpha,
\)
as well as $\kappa \defeq k / \lambda$, as in section~\ref{sec:spherical-waves}. We want to approximate the spherical waves with (scaled) plane waves in analogy with the paraxial approximation of \eqref{eq:state-at-screen-approx}. Hence, consider the symbol of a plane wave (with respect to the origin),
\begin{equation}\label{eq:plane-wave-symbol}
	\begin{split}
		\ev{e^{i\bm{k}_\pm\cdot\bhat{x}}}{\bm{z}}
		= & \ip{\bm{z}}{g(\bm{k}_\pm)\bm{z}}              \\
		= & \exp\!\left[-\frac{1}{2}\left(\norm{\bm{z}}^2
			+ \norm{g(\bm{k}_\pm)\bm{z}}^2\right)\right.      \\
			& \qquad\left.\vphantom{\frac{1}{2}}
			+ \bm{z}^\dagger (\cos\kappa +
			i\sin\kappa\, \bhat{k}_\pm\cdot\bhat\sigma)\bm{z}\right],
	\end{split}
\end{equation}
where we have invoked \eqref{eq:plane-wave-coherent-state}, together with the coherent state overlap (a special case of \eqref{eq:coherent-mel-poly}). Now $g(\bm{k}_\pm)\in\SU2$, so $\norm{g(\bm{k}_\pm)\bm{z}}^2 = \norm{\bm{z}}^2 = R$. Moreover,
\(
\bm{z}^\dagger (\bhat{k}_\pm\cdot\bhat\sigma) \bm{z}
= \bhat{k}_\pm \cdot \bm{D} / \lambda,
\)
by \eqref{eq:z-def}, and $\bhat{k}_\pm \cdot \bm{D} \approx r_\pm$, as in the commutative case, so
\begin{equation}
	\ev{e^{i\bm{k}_\pm\cdot\bhat{x}}}{\bm{z}}
	\approx e^{R(\cos\kappa - 1) + iR_\pm \sin\kappa}.
\end{equation}
This closely resembles one of the terms in \eqref{eq:pinhole-spherical-waves}, and by inspection we can see that the following state will have the desired symbol,
\begin{equation}\label{eq:nc-state-at-screen-approx}
	\qket{\psi} = \frac{M}{i}
	\left[\eta_+ \qket{\bm{k}_+}
		+ \eta_- \qket{\bm{k}_-}\right],
\end{equation}
for real dimensionless constants
\[
	\eta_\pm \defeq \frac{1}{R_\pm}\exp[(R_\pm - R)(\cos\kappa - 1)].
\]
While \eqref{eq:nc-state-at-screen-approx} and~\eqref{eq:state-at-screen-approx} appear superficially to differ in form, we should note that $\eta_\pm \approx 1 / R_\pm$ for large $L$ (or small $\kappa$). The plane waves $\qket{\bm{k}_\pm}$ each come with normalisation factors $N = \frac{1}{\sqrt{4\pi\lambda^3}}$ (as per \eqref{eq:normalisation-constant}). As in the commutative calculation, we are not much concerned with overall dimensionless normalisation factors; that said, we should include a factor of $\sqrt{\frac{\lambda}{V^{1/3}}}$ for comparison with the commutative result, given that these values define the length scales of the problem. For this reason we choose the simplest sensible overall normalisation, $M \defeq i\sqrt{\frac{\lambda}{2 V^{1/3}}}$, for the sum.

Now the probability of observing our state at $\bm{D}$ is calculated using \eqref{eq:pos-Born-rule-pure-state} as
\begin{equation}\label{eq:P-with-mel}
	\begin{split}
		P(\bm{D})
		& = \frac{1}{V^{1/3}}
		\left[\frac{\eta_+^2 + \eta_-^2}{2}
			\ev{\hat{r}}{\bm{z}} \right.                \\
			& \qquad\quad\ \left.\vphantom{\frac{\eta_-^2}{r_+^2}} + \eta_+\eta_-
			\Re\ev{e^{i\bm{k}_+\cdot\bhat{x}} \hat{r}
				e^{-i\bm{k}_-\cdot\bhat{x}}}{\bm{z}}\right].
	\end{split}
\end{equation}
The first remaining matrix element, $\ev{\hat{r}}{\bm{z}}$, is easily computed using the machinery of appendix~\ref{app:coherent-state-mels-polynomials}, specifically \eqref{eq:coherent-mel-poly}, whereby
\[
	\ev{\hat{r}}{\bm{z}} = \lambda \ev{\hat{n} + 1}{\bm{z}} = \lambda (R + 1).
\]
For the remaining matrix element --- let us call it $E$ for brevity --- we first invoke \eqref{eq:plane-wave-coherent-state} to act each of the plane waves on its adjacent coherent state (this simply produces two new coherent states, as explained in appendix~\ref{app:su2-transformations}),
\[
	E = \lambda \mel{g(-\bm{k}_+)\bm{z}}{\hat{n} + 1}{g(-\bm{k}_-)\bm{z}},
\]
and we are now again in a position to invoke \eqref{eq:coherent-mel-poly}, whereby
\begin{align*}
	E & = \lambda \exp[-\frac{1}{2}\left(\norm{g(-\bm{k}_+)\bm{z}}^2
	+ \norm{g(-\bm{k}_-)\bm{z}}^2\right) + K]                        \\
	  & \quad\times (K + 1)                                          \\
	  & = \lambda e^{K-R} (K + 1),
\end{align*}
where the constant
\(
K \defeq (g(-\bm{k}_+)\bm{z})^\dagger g(-\bm{k}_-)\bm{z}
\)
is defined as in appendix~\ref{app:coherent-state-mels-polynomials}. It now only remains to compute $K$ explicitly. Expanding its definition,
\begin{equation}\label{eq:K-matrices}
	K = \bm{z}^\dagger e^{i\lambda\bm{k}_+\cdot\bhat{\sigma}}
	e^{-i\lambda\bm{k}_-\cdot\bhat{\sigma}} \bm{z},
\end{equation}
we express the product
\(
e^{i\lambda\bm{k}_+\cdot\bhat{\sigma}}
e^{-i\lambda\bm{k}_-\cdot\bhat{\sigma}}
\)
of $\SU2$ group elements in the form
\(
e^{i\kappa_3\bhat{k}_3\cdot\bhat{\sigma}}
= \cos\kappa_3 + i\, \bhat{k}_3\cdot\bhat{\sigma}\, \sin\kappa_3,
\)
which is always possible, as explained in section~\ref{sec:plane-waves} and appendix~\ref{app:su2-transformations}. Leaving $\bm{k}_3$ and $\kappa_3$ undetermined for the moment, we can write $K$ as
\begin{equation}
	\begin{split}
		K &= R \cos\kappa_3 + i\sin\kappa_3
		\left(\bm{z}^\dagger \hat\sigma^i \bm{z}\right)[\bhat{k}_3]_i\\
		&= R\left(\cos\kappa_3+ i\,\bhat{D}\cdot\bhat{k}_3\,\sin\kappa_3\right),
	\end{split}
\end{equation}
much as we did above in \eqref{eq:plane-wave-symbol}. Finally, the BCH formula, \eqref{eq:SU2-BCH}, gives explicit formulae for $\cos\kappa_3$ and $\bhat{k}_3\sin\kappa_3$, into which we substitute
\begin{align*}
	\kappa_1   & = \kappa_2 \equiv \kappa = \lambda k,                            \\
	\bhat{k}_1 & \equiv \bhat{k}_+ = \frac{1}{\sqrt{L^2 + y_D^2 + (z_D - d)^2}}
	\begin{bmatrix}L\\ y_D\\ z_D - d\end{bmatrix},                                \\
	\bhat{k}_2 & \equiv -\bhat{k}_- = \frac{-1}{\sqrt{L^2 + y_D^2 + (z_D + d)^2}}
	\begin{bmatrix}L\\ y_D\\ z_D + d\end{bmatrix}.
\end{align*}
One of the resulting terms contains the triple product
\(
\bhat{D}\cdot(\bhat{k}_+\times\bhat{k}_-),
\)
but this vanishes since the three vectors in question are co-planar, as depicted in figure~\ref{fig:pinhole-planar}. We are left with
\begin{equation}
	\begin{split}
		K &= R\cos^2\kappa + R\sin^2\kappa\cos(\alpha+\beta) \\
		&\quad + i R\sin\kappa\cos\kappa (\cos\alpha - \cos\beta),
	\end{split}
\end{equation}
where we have written each scalar product appearing in \eqref{eq:SU2-BCH} in terms of the angles $\alpha$ and $\beta$ from figure~\ref{fig:pinhole-planar}. For brevity, let us introduce two more constants,
\begin{equation}
	\begin{split}
		A & \defeq \Re K
		= R \left(\cos^2\kappa + \cos(\alpha + \beta)\sin^2\kappa\right), \\
		B & \defeq \Im K
		= R\sin\kappa\cos\kappa (\cos\alpha - \cos\beta).
	\end{split}
\end{equation}

Finally, inserting everything into \eqref{eq:P-with-mel}, we obtain our probability distribution,
\begin{equation}\label{eq:P}
	\begin{split}
		P(\bm{D})
		&= \frac{\lambda}{V^{1/3}}
		\left[\frac{\eta_+^2 + \eta_-^2}{2} (R + 1) \right. \\
			&\quad\left.\vphantom{\frac{\eta_-^2}{r_+^2}}
			+ \eta_+\eta_- e^{A - R} ((A + 1)\cos B - B\sin B) \right].
	\end{split}
\end{equation}

\section{Discussion}\label{ch:discussion}

In this section we discuss our result, \eqref{eq:P}, and consider various limiting cases. Specifically, we confirm that in the commutative limit, $\lambda\to 0$, \eqref{eq:P} reduces to the commutative result, \eqref{eq:P-commutative}. Following that, we consider the classical limit, deriving the specific conditions under which we observe a quantum-to-classical transition, and finally we derive how \eqref{eq:P} is affected by allowing and entire collection of $N$ particles to pass through the pinhole setup at once.

\subsection{Qualitative discussion of form}

It is worth first making a few cursory remarks on the key features of our distribution. Qualitatively, the individual sub-expressions of \eqref{eq:P} perform the following functions,
\begin{equation}\label{eq:P-annotated}
	\begin{split}
		P(\bm{D}) &\sim
		\smash[b]{\color{gray}\underbrace{\color{black}
			\frac{\eta_+^2 + \eta_-^2}{2} (R + 1)
		}_{\color{black}\substack{\text{bimodal shaping}\\\text{function}}}}\ +\
		\smash[b]{\color{gray}\underbrace{\color{black}\vphantom{\frac{r^2_+}{2r_+}}
				\eta_+\eta_- e^{A - R}
			}_{\color{black}\substack{\text{interference}\\\text{suppression}}}}
		\\[2.5em]&\quad\times\
		\smash[b]{\color{gray}\underbrace{\color{black}\vphantom{\frac{r^2_+}{2r_+}}
		((A + 1)\cos B - B\sin B)
		}_{\color{black}\substack{\text{interference terms}}}}.
		\vphantom{\underbrace{A_1}
		_{\substack{\text{s}\\\text{f}}}}
	\end{split}
\end{equation}
Notably, $e^{A - R}$ indeed always acts to suppress the interference, given that $A - R = R\sin^2\kappa\, (\cos(\alpha + \beta) - 1)$ is manifestly always nonpostive, since $\kappa\in[0,\pi)$.

\begin{figure*}[t]
	\centering
	\begin{subfigure}[b]{0.45\textwidth}
		\centering
		\includegraphics[width=\textwidth]{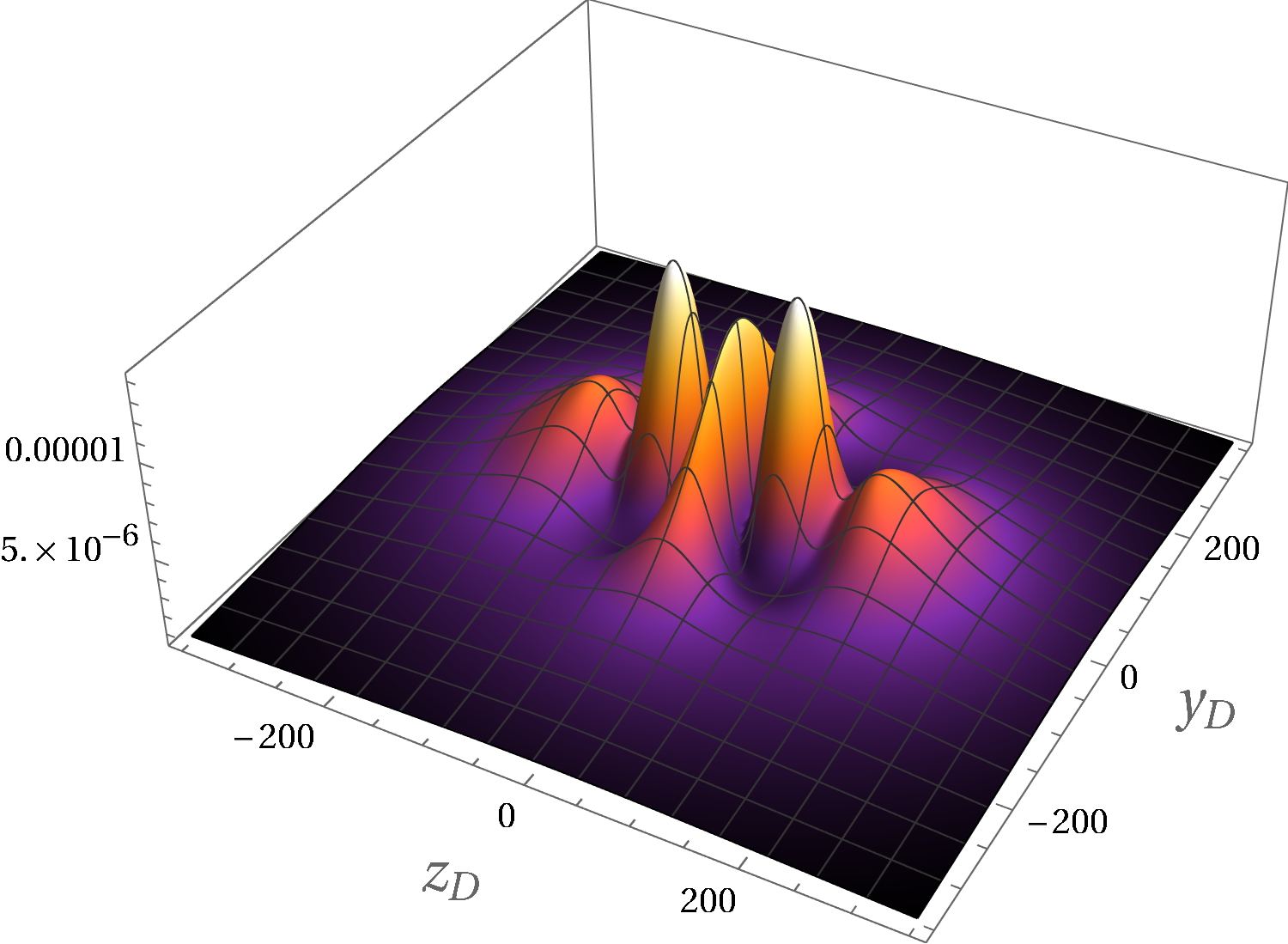}

		\vspace{0.5em}
		\includegraphics[width=\textwidth]{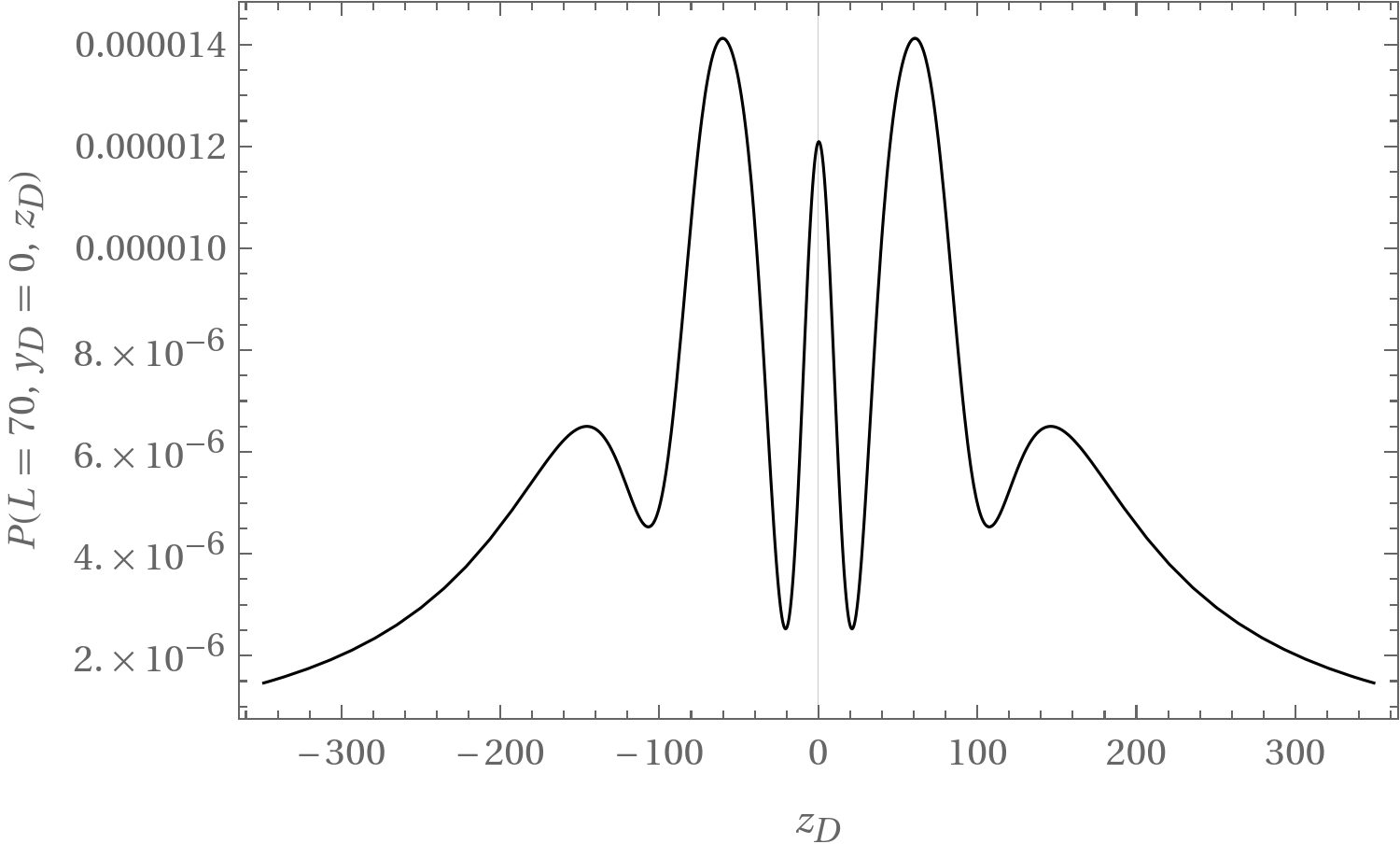}
		\caption{$k=0.22$}
		\label{fig:Ld70-lmda.1:k.22}
	\end{subfigure}
	\hfill
	\unskip\ \color{gray}{\vrule}\
	\begin{subfigure}[b]{0.45\textwidth}
		\centering
		\includegraphics[width=\textwidth]{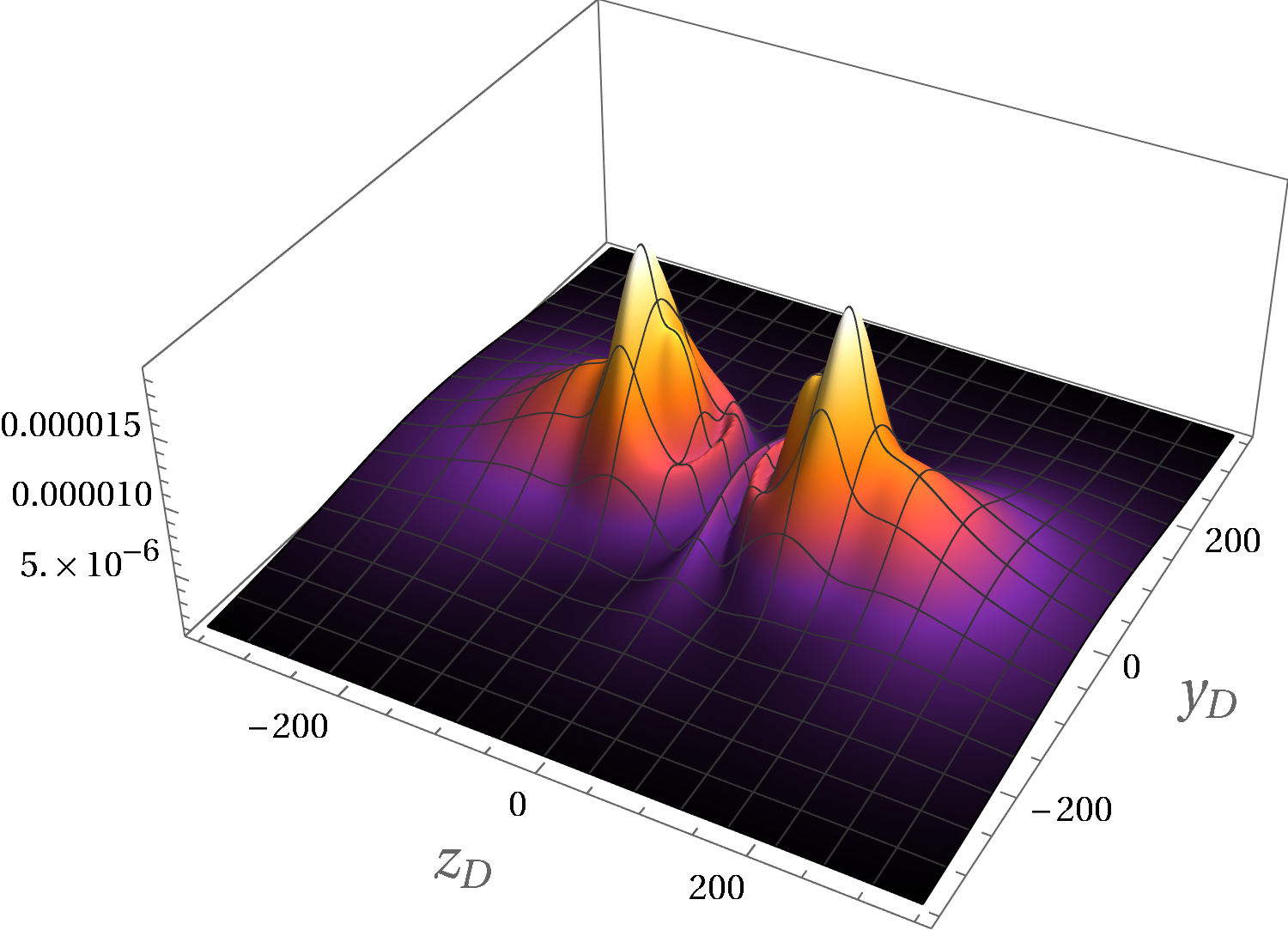}

		\vspace{0.5em}
		\includegraphics[width=\textwidth]{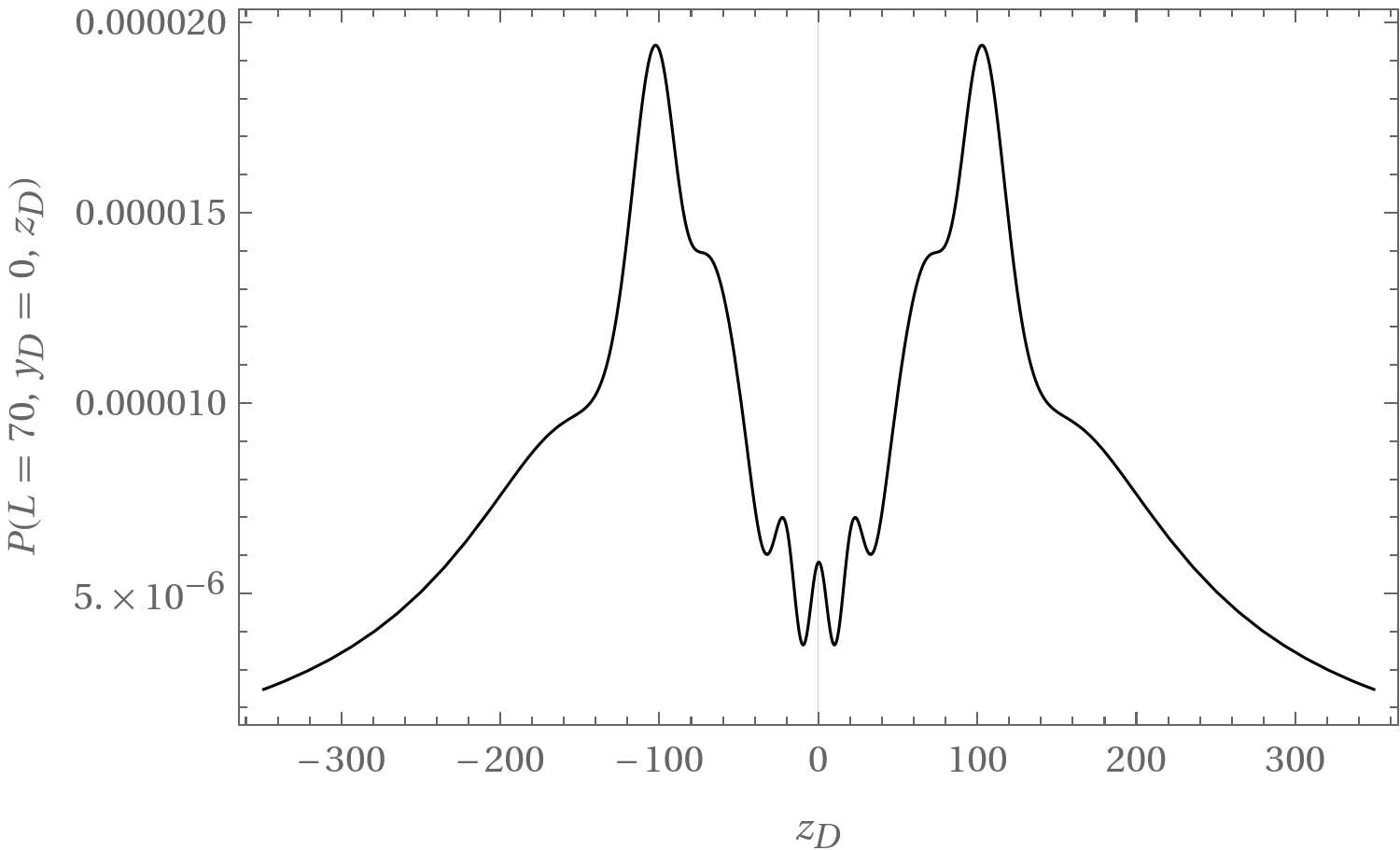}
		\caption{$k=0.44$}
		\label{fig:Ld70-lmda.1:k.44}
	\end{subfigure}

	\caption{
		Surface plots of $\frac{1}{4\pi\lambda^2 r}P(\bm{D})$, with $y_D=0$ 	traces, for different values of $k$.
		Other parameters are $L=d=70,\lambda=0.1$. All lengths are in arbitrary units.
	}
	\label{fig:Ld70-lmda.1}
\end{figure*}

For understanding qualitative features, it is also helpful to examine a plot. Figure~\ref{fig:Ld70-lmda.1} shows surface plots of \eqref{eq:P} for some arbitrarily-chosen parameter values, along with the $y_D=0$ traces. Instead of $P(\bm{D})$, we actually plot $\frac{1}{4\pi\lambda^2 r}P(\bm{D})$, since this is the spatial probability density corresponding to $P(\bm{D})$, as explained in section~\ref{sec:pos-measurement}; this helps with later qualitative comparison with the commutative limit. The plots clearly exhibit the expected bimodal distribution superimposed with interference. Moreover, the interference is evidently stronger for one set of parameters than for the other, and increased suppression is correlated with increased localisation of the distribution.

More importantly, the plots also suggest that the distribution is reflection-symmetric. Indeed, we can easily algebraically verify its symmetry under $z_D \longleftrightarrow -z_D$ --- each of the following quantities are manifestly invariant under this replacement: $R$; $\cos(\alpha+\beta)$, and hence $e^A$; $\cos B$ (as $B$ accrues a minus sign under the reflection); and finally $\eta_+\eta_-$ and $\eta_\pm^2$. This would hardly be remarkable, were it not for the fact that the double-slit interference pattern in the 2D Moyal plane is \emph{asymmetric} under reflection~\cite{pittaway2021quantum}. The restoration of reflection symmetry in our analogous setup in 3D fuzzy space confirms the expectations of~\cite{pittaway2021quantum} that the reflection-asymmetry in the Moyal plane interference pattern arises only because the Heisenberg-Moyal commutation relations break \emph{rotational} symmetry (as explained in section~\ref{sec:formalism}). The distribution is likewise (unsurprisingly) symmetric under the reflection $y_D \longleftrightarrow -y_D$, and thus also under $180^\circ$ rotation about the $x$-axis (the composition of these two reflections).

\subsection{Commutative limit}\label{sec:commutative-lim}

We now take the commutative limit $\lambda\to0^+$, and show that \eqref{eq:P} indeed reproduces \eqref{eq:P-commutative}, as it must. There is actually a small subtlety which we have already foreshadowed. That is, since $P_\text{comm}(\bm{D})$ represents a \emph{spatial} density (with dimensions of length$^{-3}$), and $P(\bm{D})$ does not (being dimensionless), we should instead expect to obtain $P_\text{comm}(\bm{D})$ as the limit
\[
	\lim_{\lambda\to 0^+} \frac{1}{4\pi r\lambda^2} P(\bm{D}),
\]
rather than simply $\lim_{\lambda\to 0^+} P(\bm{D})$, since (as shown in section~\ref{sec:pos-measurement}) $\frac{1}{4\pi r\lambda^2} P(\bm{D})$ is precisely the spatial density corresponding to $P(\bm{D})$.

Now, as we send $\lambda\to 0^+$, $B$ clearly obtains a finite non-zero limit, which we call $B_0$, whereas $A$ diverges like $R\equiv r/\lambda$. But the divergence of $A$ is cancelled in the expression $A-R$, which altogether vanishes, meaning the interference suppression vanishes in this limit, $e^{A-R} \to 1$. Finally, the numerator, $R_\pm\eta_\pm$, of $\eta_\pm$ tends to $1$, so that $\eta_\pm / \lambda \to 1 / r_\pm$. Using the above results (invoking continuity wherever applicable), we compute the commutative limit
\begin{align*}
	 & \lim_{\lambda\to 0} \frac{1}{4\pi r\lambda^2} P(\bm{D})     \\
	 & = \frac{1}{4\pi V^{1/3}\, r_+r_-}
	\left[\frac{r_+^2 + r_-^2}{2r_+r_-}\right.                     \\*
	 & \qquad\qquad\qquad\quad\ \left.\vphantom{\frac{r_-^2}{r_-}}
	+ \cos(rk (\cos\alpha - \cos\beta))\right]                     \\
	 & = P_\text{comm}(\bm{D}),
\end{align*}
as required.

\begin{figure*}[t]
	\centering
	\begin{subfigure}[b]{0.45\textwidth}
		\centering
		\includegraphics[width=\textwidth]{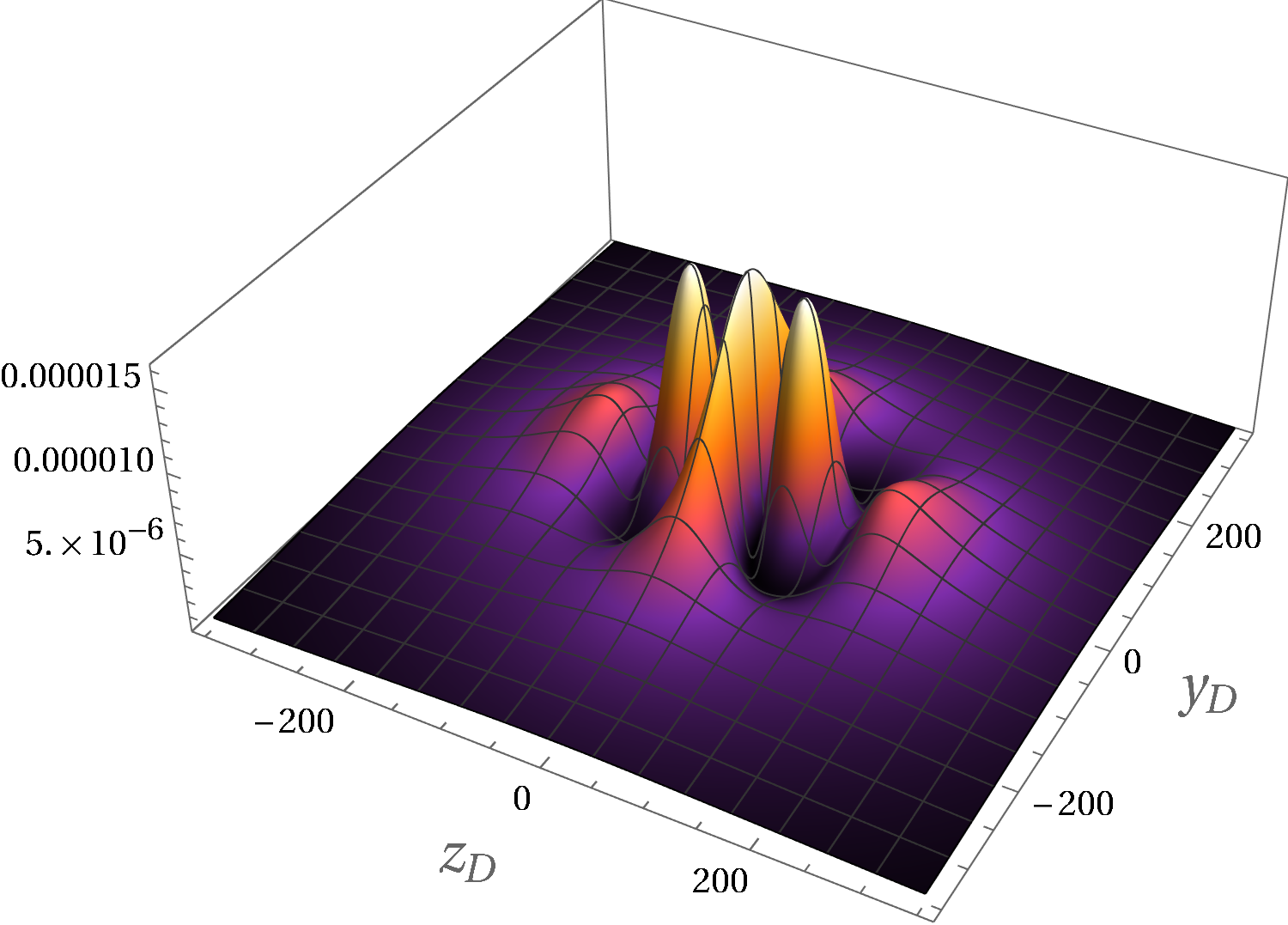}

		\vspace{0.5em}
		\includegraphics[width=\textwidth]{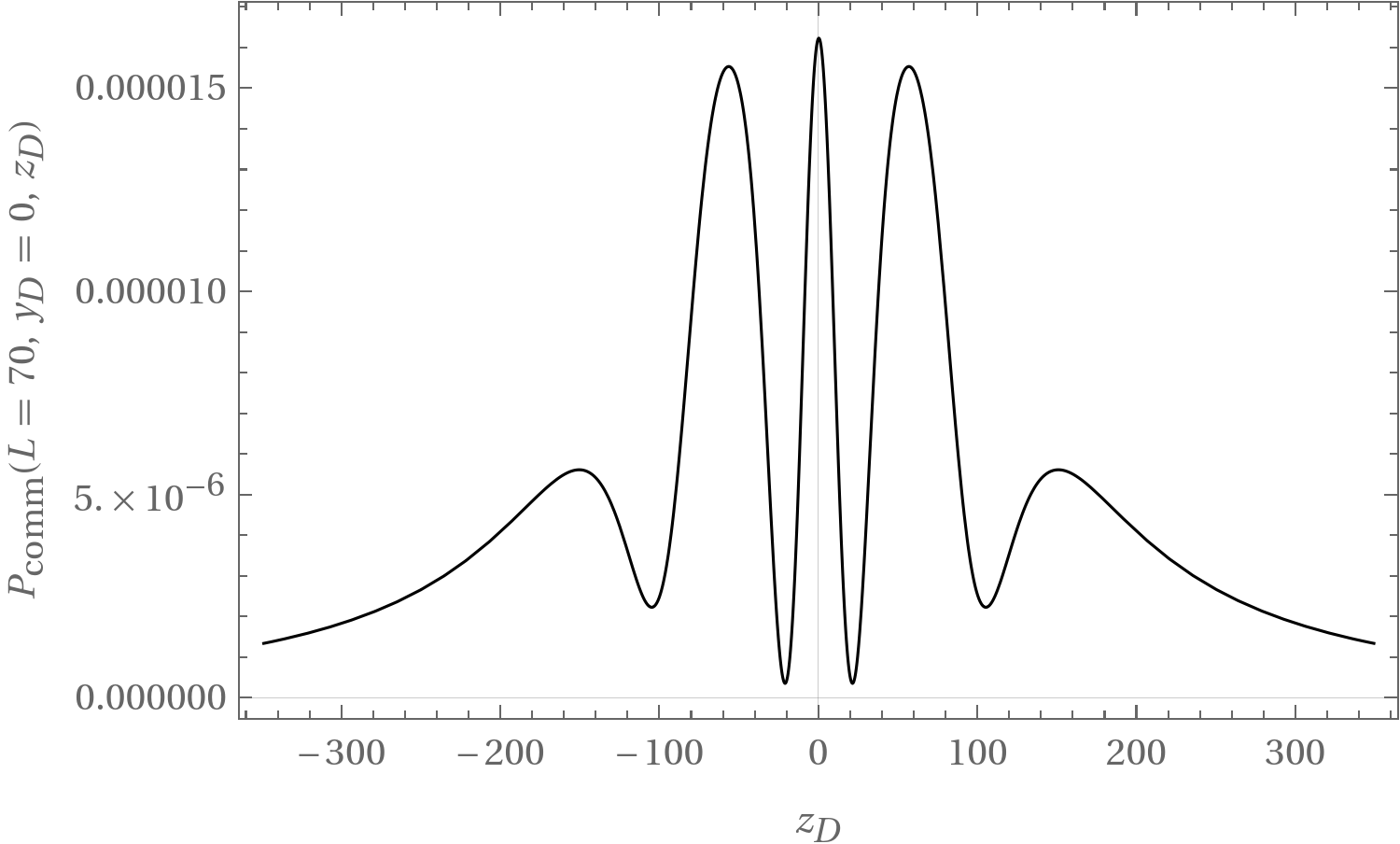}
		\caption{$k=0.22$}
		\label{fig:c-Ld70-lmda.1:k.22}
	\end{subfigure}
	\hfill
	\unskip\ \color{gray}{\vrule}\
	\begin{subfigure}[b]{0.45\textwidth}
		\centering
		\includegraphics[width=\textwidth]{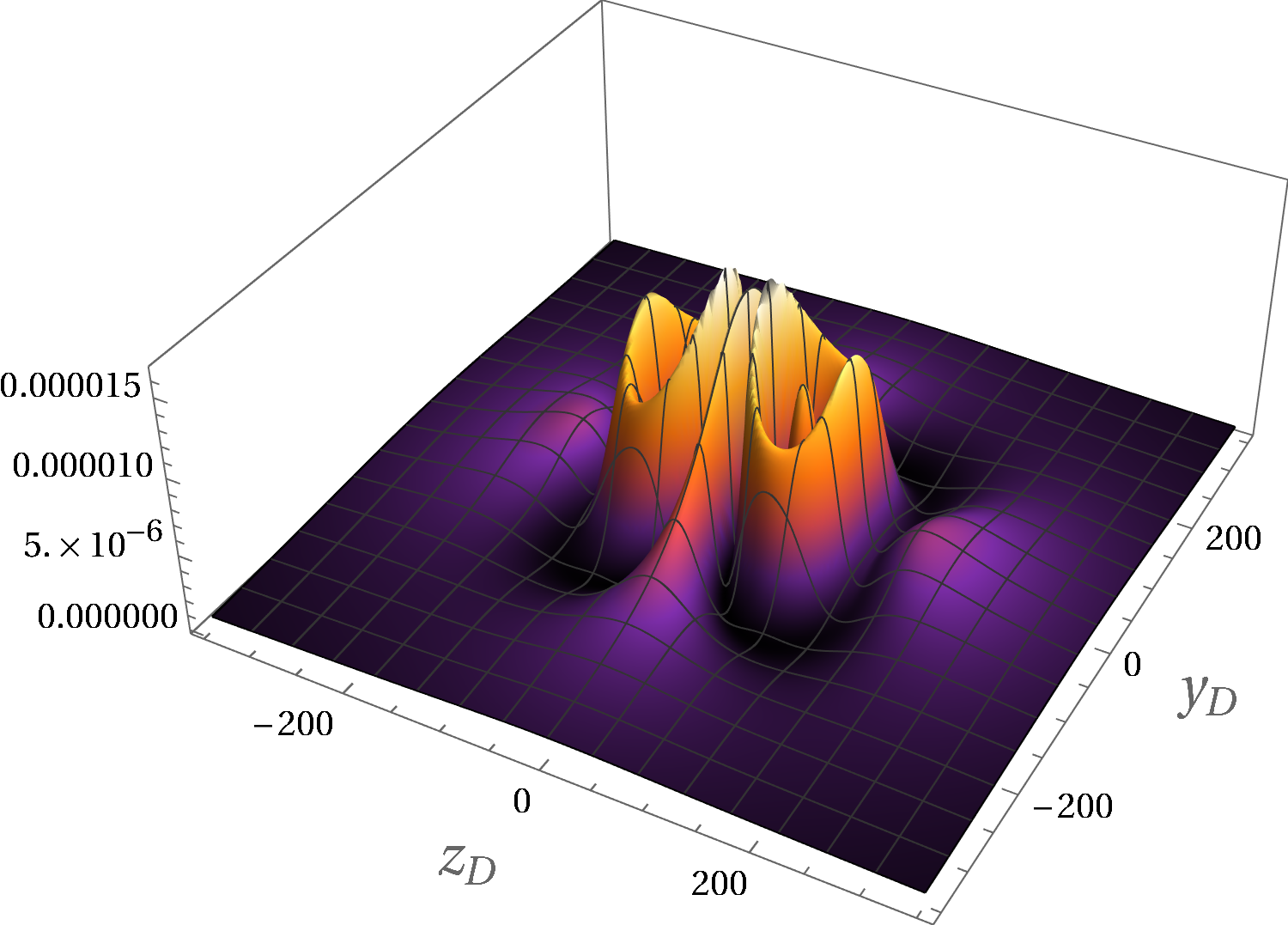}

		\vspace{0.5em}
		\includegraphics[width=\textwidth]{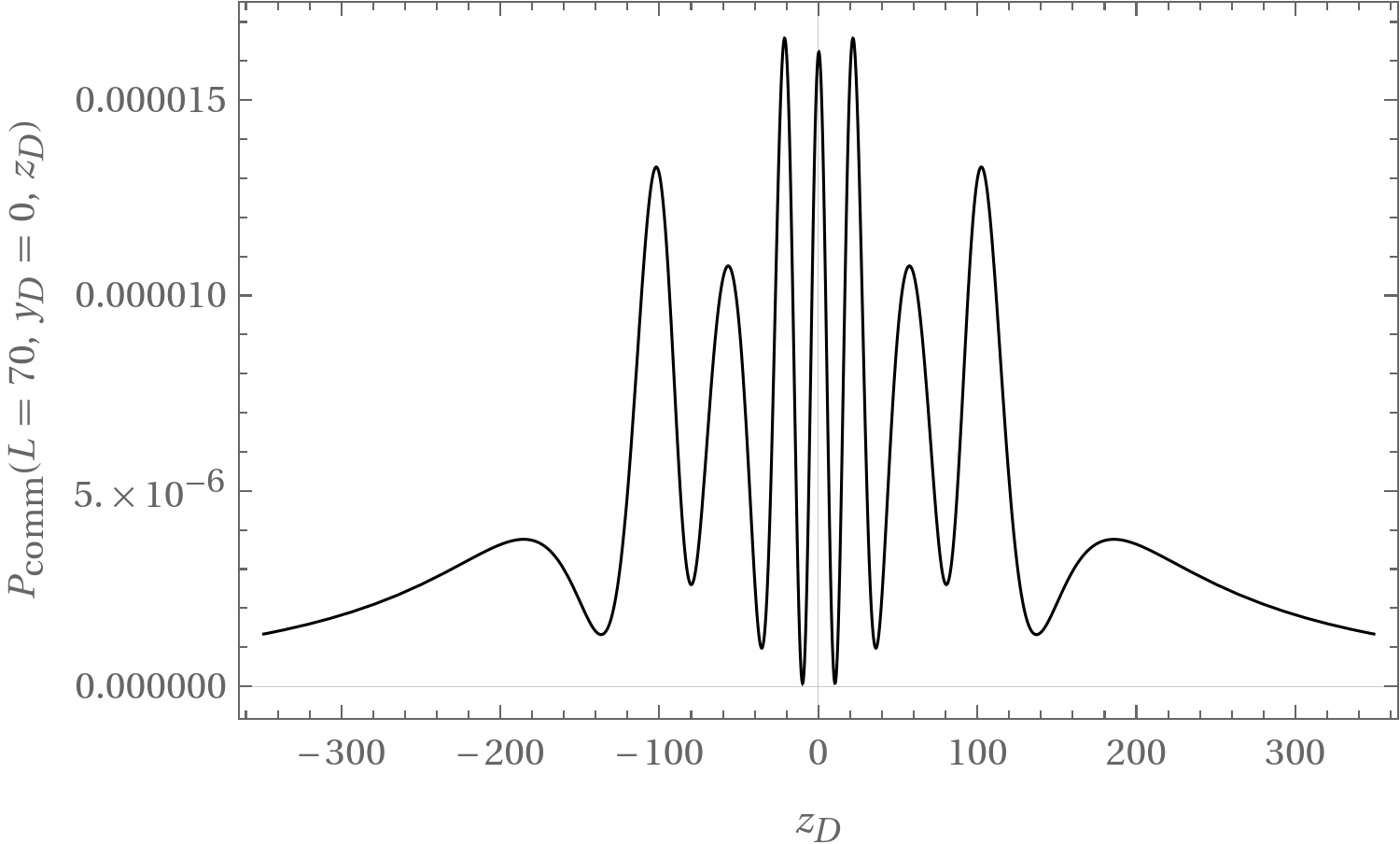}
		\caption{$k=0.44$}
		\label{fig:c-Ld70-lmda.1:k.44}
	\end{subfigure}

	\caption{
		Surface plots of $P_{\text{comm}}(\bm{D})$, with $y_D=0$ traces, for the same values of $k$, $L$, and $d$ as in figure~\ref{fig:Ld70-lmda.1}.
	}
	\label{fig:c-Ld70-lmda.1}
\end{figure*}

Having established that we have the correct commutative limit, we pause to compare its behaviour qualitatively with that of $P(\bm{D})$. To this end, we plot $P_\text{comm}(\bm{D})$ in figure~\ref{fig:c-Ld70-lmda.1} for the same parameter values (except $\lambda$, of course) as in figure~\ref{fig:Ld70-lmda.1}.
Comparing figures~\ref{fig:c-Ld70-lmda.1} and~\ref{fig:Ld70-lmda.1}, the low-momentum ($k=0.22$) distributions are unsurprisingly similar, but where in the non-commutative distribution of figure~\ref{fig:Ld70-lmda.1} we observe a \emph{suppression} of interference for higher momentum ($k=0.44$), the commutative distribution of figure~\ref{fig:c-Ld70-lmda.1} actually exhibits completely the opposite behaviour, showing more pronounced interference at this higher $k$. This shows that the momentum-dependent quantum-to-classical transition that we observe is a uniquely non-commutative phenomenon. In the next section, we will consider this transition more carefully.

\subsection[Classical limit]{Classical behaviour and quantum-to-classical transition}
\label{sec:classical-lim}

\begin{figure}[b]
	\centering
	\includegraphics[width=0.43\textwidth]{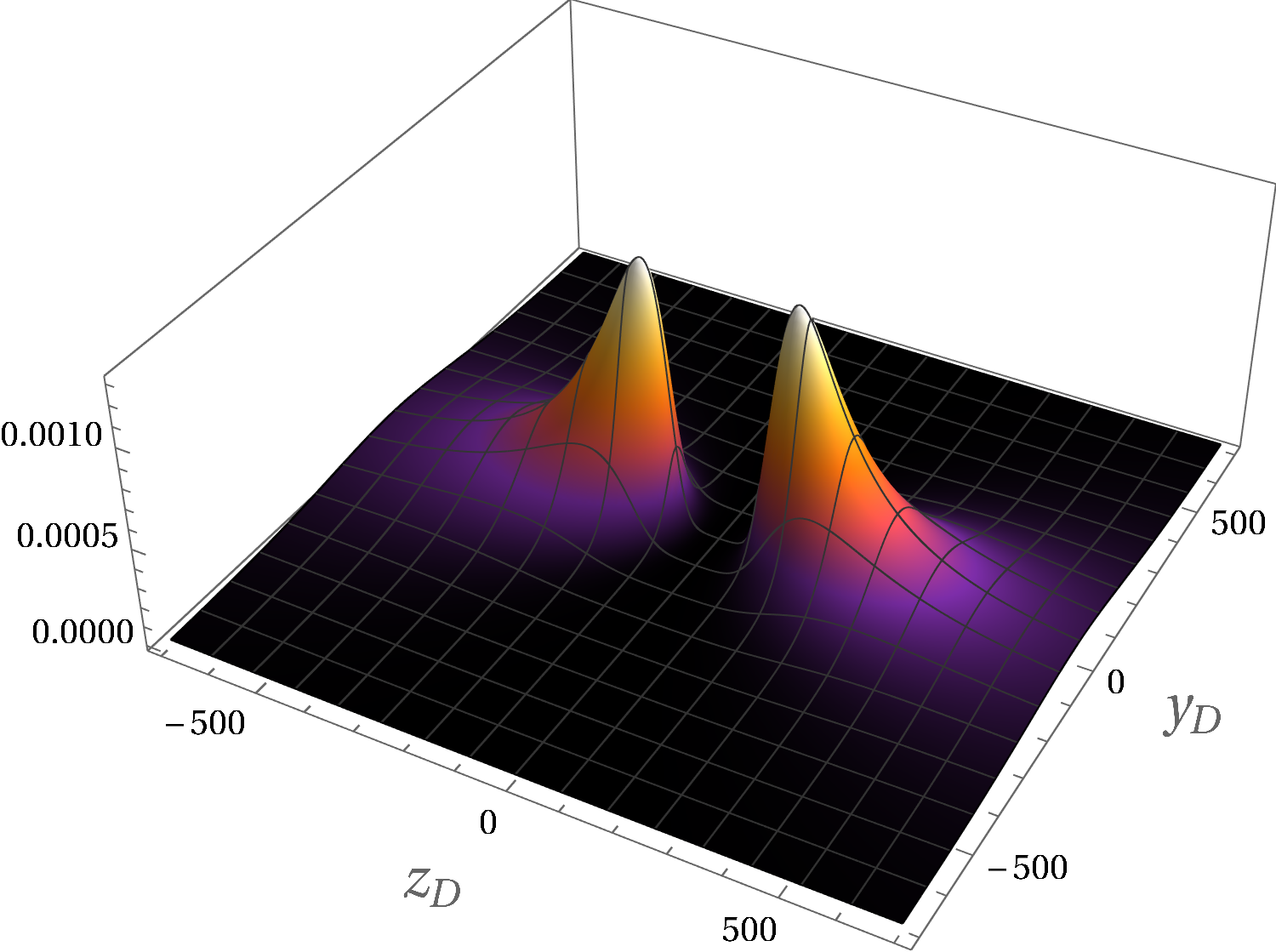}
	\caption{
		Surface plot of $P(\bm{D})$ with $k=1$ (and $L$, $d$ and $\lambda$ as in figure~\ref{fig:Ld70-lmda.1}), exhibiting classical-regime behaviour of being bimodal and localised with no visible interference.
	}
	\label{fig:cl-Ld70-lmda.1-k1}
\end{figure}

We recognise the ``classical-regime'' of our distribution to consist of any parameter combinations that result in strong interference suppression, leaving behind the underlying bimodal distribution that one would expect to emerge classically from our pinhole setup. An example of a distribution in this regime is shown in figure~\ref{fig:cl-Ld70-lmda.1-k1}.
Looking at figure~\ref{fig:Ld70-lmda.1}, or even the double slit interference in the Moyal plane~\cite{pittaway2021quantum}, we might expect this regime to emerge in a large momentum limit. However, recall that, like the plane wave energy of \eqref{eq:plane-wave-energy}, the probability distribution of \eqref{eq:P} is periodic in $k$ with period $2\pi/\lambda$; such periodicity does not occur in the Moyal plane~\cite{pittaway2021quantum}. The upshot is that we do not obtain a well-defined limit by sending $k\to\infty$, so it is not apparent what a large momentum limit should entail. Indeed, we show that suppression can occur even for arbitrarily small momenta.

Recall from \eqref{eq:P-annotated} that the factor responsible for interference suppression is
\[
	e^{A-R} = \exp[R\sin^2\kappa(\cos(\alpha + \beta) - 1)].
\]
First we write $\cos(\alpha + \beta)$ in terms of the length scales in the problem. Applying the cosine rule to figure~\ref{fig:pinhole-planar}, we obtain
\begin{align*}
	\cos(\alpha + \beta)
	= \frac{r^2 - d^2}{\sqrt{(r^2 + d^2)^2 - 4z_D^2 d^2}}
	\approx 1 - \frac{2d^2}{r^2},
\end{align*}
to first order in each of $d/r \ll1$ and $z_D/r \ll1$. Suppose for the moment that $\lambda k / 2 \ll 1$, as would be the case at low momentum. Expanding the exponent,
\begin{align*}
	A - R & = \lambda r k^2(\cos(\alpha + \beta) - 1) + \mathcal{O}(\lambda^3 k^3) \\
	      & \approx -\frac{2\lambda k^2 d^2}{r}
\end{align*}
to leading order, we identify the condition for strong suppression as $2\lambda k^2 d^2 / r \gg 1$. It is helpful to rewrite this condition in terms of energy, and given our assumption on $\lambda k$, we can expand~\eqref{eq:plane-wave-energy} as
\(
E \approx \frac{\hbar^2 k^2}{2m}.
\)
The condition for suppression then becomes
\begin{equation}\label{eq:classical-condition}
	\frac{4\lambda d^2 m E}{r \hbar^2} \gg 1
	\quad\iff\quad r \ll \frac{4\lambda d^2 m E}{\hbar^2},
\end{equation}
so that, in particular, suppression is only visible at sufficiently small distances. Combining this with the large-separation assumption, we note that this particular expression is valid in the regime
\begin{equation}\label{eq:classical-condition-constraint}
	1 \ll \frac{r}{d} \ll \frac{4\lambda d m E}{\hbar^2}.
\end{equation}
Note that in the commutative limit $\lambda=0$, the upper limit on the distance of observation vanishes and interference is prevalent at all length scales of observation.

There are two remarkable features of~\eqref{eq:classical-condition} that distinguish it from the analogous classicality condition for double-slit interference in the Moyal plane~\cite{pittaway2021quantum}. Firstly, our suppression strength is here also affected by the distance $r$ at which measurement is performed. Secondly (indeed, consequently), here we are theoretically capable of observing suppression even at low momentum (by taking $r$ sufficiently small). These are important testable predictions of our three-dimensional theory.

At this point, we can perform a similar back-of-the-envelope estimate to that in~\cite{pittaway2021quantum} to get a sense of when we should expect to see suppression in practice. Assuming $\lambda$ to be of the order of a Planck length, and $d$ to be on the order of $1\SI{}{\centi\meter}$, and considering an electron of energy $1\SI{}{\electronvolt}$, we expect to observe suppression of the interference when
\begin{align*}
	r & \lesssim \frac{(10^{-35}\, \SI{}{\meter})
		(10^{-2}\, \SI{}{\meter})^2 (10^{-31}\, \SI{}{\kilogram})
		(10^{-19}\, \SI{}{\kilogram\square\meter\per\square\second})}
	{(10^{-34}\, \SI{}{\kilogram\square\meter\per\second})^2} \\
	  & = 10^{-21}\, \SI{}{\meter}.
\end{align*}
We should of course note that the values used here fall well within the constraints, $\lambda k \ll 1$ and \eqref{eq:classical-condition-constraint}, for which \eqref{eq:classical-condition} is valid. The required value of $r$ is obviously extremely small, so we would not detect any suppression at these energies. Conversely, to observe suppression at lengths on the order of a meter requires an electron to have energies around $10^{19}\, \SI{}{\electronvolt}$ --- again, quite undetectable. Thankfully this is not concerning, since in the next section we will show how, once multiple particles are allowed to interfere, the suppression becomes much more pronounced.

\subsection[Macroscopic behaviour]{Macroscopic behaviour}\label{sec:macroscopic-lim}

In this section we derive how \eqref{eq:P} scales with increased particle number. Specifically, we modify our interference setup to consider a collection of particles passing through the pinholes. As usual our interest is in the collective dynamics of this collection of particles, \textit{i.e.}~the centre-of-mass dynamics, which we expect to exhibit classical behaviour for a sufficiently large number of particles.

We begin by straightforwardly extending our core definitions to the multi-particle case. Consider a macroscopic object comprised of $N$ particles of equal mass $m$, with total mass
\(
M \defeq Nm.
\)
The total system is described by the Hilbert space, $\Hq^\text{tot}$, constructed from particle Hilbert spaces, $\Hq^{(n)}$, in the usual way,
\[
	\Hq^{\text{tot}} \defeq \bigotimes_{n=1}^N \Hq^{(n)}.
\]
For simplicity, we neglect the symmetrisation (resp.~antisymmetrisation) required by boson (resp.~fermion) statistics. Let the $n^\text{th}$ particle coordinate operators be denoted $\hat{x}_i^{(n)}$; we assume that coordinate operators belonging to different particles commute,
\begin{equation}
	\left[\hat{x}_i^{(l)}, \hat{x}_j^{(n)}\right]
	= 2i\lambda\epsilon_{ijk}\delta_{ln} \hat{x}_k^{(l)}.
\end{equation}
The collective motion is described with \emph{centre-of-mass} coordinates,
\begin{equation}
	\bhat{x}^{(\text{CM})} \defeq \frac{1}{N} \sum_{n=1}^N \bhat{x}^{(n)},
\end{equation}
satisfying the commutation relations
\begin{equation}
	\left[\hat{x}_i^{(\text{CM})}, \hat{x}_j^{(\text{CM})}\right]
	= 2i\frac{\lambda}{N}\epsilon_{ijk}\, \hat{x}_k^{(\text{CM})},
\end{equation}
which are manifestly identical to those of $\hat{x}_i$, only with
\(
\tilde\lambda \equiv \frac{\lambda}{N}
\)
in place of $\lambda$. As usual, we also have the \emph{relative} coordinates,
\(
\bhat{\xi}^{(n)} \defeq \bhat{x}^{(n)} - \bhat{x}^{(\text{CM})},
\)
which are easily seen to have vanishing sum,
\[
	\sum_{n=0}^N \bhat{\xi}^{(n)} = \bm{0}.
\]

At this point, the normal procedure is to transform to centre-of-mass and relative coordinates in the Hamiltonian. Assuming translational invariance, \textit{i.e.}~interactions depending only on the relative coordinates, allows one to decouple the centre-of-mass and relative motion, leaving free-particle centre-of-mass dynamics, with all internal dynamics captured by the relative dynamics.

Here the procedure is more involved for two reasons: firstly the more complicated form of the Laplacian operator, and secondly the fact that the centre-of-mass and relative coordinates no longer commute, rendering the clean decoupling of centre-of-mass and relative dynamics impossible. Even so, it is still possible to isolate the centre-of-mass dynamics, as we proceed to show. The first point to realise is that the centre-of-mass dynamics are, as usual, governed by the free particle contribution to the Hamiltonian, which reads
\[
	\hat{H}^{\text{tot}} = \sum_{n=1}^N \hat{H}^{(n)},
\]
where each $\hat{H}^{(n)}$ is a straightforward generalisation of \eqref{eq:free-schrodinger},
\[
	\hat{H}^{(n)} \defeq -\frac{\hbar^2}{2m} \hat{\Delta}^{(n)}.
\]
Indeed, wherever we use a superscript $(n)$ on a known operator, it should be assumed to act on the $n^{\text{th}}$ Hilbert space, $\Hq^{(n)}$, but otherwise be defined as usual. Now the (plane wave) eigenstates of $\hat{H}^{\text{tot}}$ are (up to normalisation)
\begin{equation*}
	\qket{\bm{k}^{(i\cdots N)}}
	\equiv \qket{\bm{k}^{(1)},\dots,\bm{k}^{(N)}}
	\defeq \exp[i\sum_{n=1}^N \bm{k}^{(n)} \cdot \bhat{x}^{(n)}],
\end{equation*}
with eigenvalues
\begin{equation}\label{eq:N-particle-plane-wave-energy}
	\hat{H}^{\text{tot}} \qket{\bm{k}^{(i\cdots N)}}
	= \left[\frac{2\hbar^2}{m\lambda^2} \sum_{n=1}^N
		\sin^2\left(\frac{k^{(n)}\lambda}{2}\right)\right]
	\qket{\bm{k}^{(i\cdots N)}};
\end{equation}
this is a simple extension of the single particle plane waves, as per section~\ref{sec:plane-waves}. For one of these plane wave solutions, we may define the \emph{total momentum},
\[
	\bm{k}^{\text{tot}} \defeq \sum_{n=1}^N \bm{k}^{(n)},
\]
as well as the \emph{relative momenta},
\[
	\bm{q}^{(n)} \defeq \bm{k}^{(n)} - \frac{1}{N} \bm{k}^{\text{tot}},
\]
noting, as with relative coordinates, the vanishing sum
\begin{equation}\label{eq:rel-momentum-sum}
	\sum_{n=1}^N \bm{q}^{(n)} = \bm{0}.
\end{equation}
Rewriting the $N$-particle plane wave in terms of total and relative momenta yields
\begin{align*}
	\qket{\bm{k}^{(i\cdots N)}}
	= \exp[i\bm{k}^{\text{tot}} \cdot \bhat{x}^{(\text{CM})}
	+ i\sum_{n=1}^N \bm{q}^{(n)} \cdot \bhat{x}^{(n)}].
\end{align*}
For non-interacting particles, one expects the energy to have two contributions --- the centre-of-mass energy and that of the relative (internal) motion. If all of the relative momenta vanish, one therefore expects the latter energy contribution to vanish, enabling one to isolate the centre-of-mass dynamics. Let us assume therefore that $\bm{q}^{(n)} = \bm0$ for all $n\in\{1,2,\dots,N\}$. In this case, the plane wave eigenstates are especially simple,
\begin{equation}\label{eq:CM-plane-wave}
	\qket{\bm{k}^{(i\cdots N)}}	= \exp[i\bm{k}^{\text{tot}} \cdot \bhat{x}^{(\text{CM})}],
\end{equation}
as are the corresponding energy eigenvalues,
\begin{equation}\label{eq:CM-plane-wave-energy}
	\hat{H}^{\text{tot}} \qket{\bm{k}^{(i\cdots N)}}
	= \frac{2\hbar^2}{M\tilde{\lambda}^2}
	\sin^2\left(\frac{k^{\text{tot}}\tilde\lambda}{2}\right)
	\qket{\bm{k}^{(i\cdots N)}}.
\end{equation}
With no relative motion, we can also simplify the form of $\hat{H}^\text{tot}$. To see this, first define the total boson number operator in the obvious way,
\[
	\hat{n}^\text{tot} \defeq \sum_{n=1}^N \hat{n}^{(n)},
\]
then split the Hamiltonian as $\hat{H}^\text{tot} \equiv \hat{H}_0 + \hat{H}_1$, where the action of each term on a state $\psi\in\Hq^{\text{tot}}$ (temporarily suppress ket notation for simplicity) is defined by
\begin{equation}\label{eq:Hamiltonian-pieces}
	\begin{split}
		&\hat{H}^\text{tot}\, \psi
		= \frac{\hbar^2}{2m\lambda^2} \sum_{n=1}^N \frac{1}{\hat{n}^{(n)} + 1}
		[{a_\alpha^{(n)}}^{\!\dagger}\!,[a_\alpha^{(n)},\psi]] \\
		&= {\color{gray}\underbrace{\color{black}
		\frac{\hbar^2}{2M\tilde{\lambda}^2}\, \frac{1}{\hat{n}^\text{tot} + N}
		\sum_{n=1}^N [{a_\alpha^{(n)}}^{\!\dagger}\!,[a_\alpha^{(n)},\psi]]
		}_{\color{black} \hat{H}_0\, \psi}}\\
		&\ + {\color{gray}\underbrace{\color{black}
			\frac{\hbar^2}{2m\lambda^2} \sum_{n=1}^N
			\left(\frac{1}{\hat{n}^{(n)} + 1} - \frac{N}{\hat{n}^\text{tot} + N}\right)
			[{a_\alpha^{(n)}}^{\!\dagger}\!,[a_\alpha^{(n)},\psi]]
		}_{\color{black} \hat{H}_1\, \psi}}.
	\end{split}
\end{equation}
Next, note that only the first term, $\hat{H}_0$, contributes to the energy eigenvalue of
\(
\psi_\text{CM} \equiv \qket{\bm{k}^{(i\cdots N)}}.
\)
Indeed, we find that
\[
	\left[{a_\alpha^{(n)}}^{\!\dagger}\!, \left[a_\alpha^{(n)}, \psi_\text{CM}\right]\right]
	= \frac{4\hat{r}^{(n)}}{\lambda}
	\sin^2\left(\frac{k^\text{tot}\lambda}{2N}\right) \psi_\text{CM},
\]
using the same argument as in the derivation of \eqref{eq:plane-wave-energy} presented in section~\ref{sec:plane-waves}, only with the replacement
\(
\bm{k} \to \bm{k}^{\text{tot}} / N
\)
and the relevant $(n)$ superscripts. Then the action of $\hat{H}_0$ on $\psi_\text{CM}$,
\begin{align*}
	\hat{H}_0\, \psi_\text{CM} = \frac{2\hbar^2}{M\tilde{\lambda}^2}
	\sin^2\left(\frac{k^\text{tot}\tilde\lambda}{2}\right)
	\psi_\text{CM},
\end{align*}
already accounts for the full energy, as per \eqref{eq:CM-plane-wave-energy}, so
\(
\hat{H}_1 \qket{\bm{k}^{(i\cdots N)}} = 0,
\)
meaning that, for plane wave solutions without relative motion, the free Hamiltonian, $\hat{H}^\text{tot}$, reduces to just the first term, $\hat{H}_0$. The conclusion is that $\hat{H}_0$ comprises the part of $\hat{H}^\text{tot}$ responsible for centre-of-mass dynamics, whereas $\hat{H}_1$ constitutes the part responsible for internal dynamics and the coupling between internal- and centre-of-mass dynamics.

For the more general case where the $\bm{q}^{(n)}$ are allowed to be non-zero, there is a free-particle (\textit{i.e.}~quadratic in $q^{(n)}$) contribution to the energy arising from the internal motion, as well as higher-order corrections in $\lambda$ reflecting the coupling between centre-of-mass and internal dynamics, as can be verified directly by expanding the right-hand-side of \eqref{eq:N-particle-plane-wave-energy} in the $q^{(n)}$ and using \eqref{eq:rel-momentum-sum}. Even if interactions depending only on relative coordinates are introduced, the Hamiltonian can still be split as in \eqref{eq:Hamiltonian-pieces}, with the interactions forming part of $\hat{H}_1$. In this case, the interactions also contribute to the internal energy as a binding energy, and to the coupling of the centre-of-mass and internal dynamics. Once again, the latter is of higher-order in $\lambda$, and we can therefore decouple the centre-of-mass and internal dynamics to lowest order in $\lambda$ and include higher-order effects perturbatively. Finally, we remark that, as the internal energy is (to leading order in $\lambda$) quadratic in the $q^{(n)}$, we can apply the equipartition theorem, which implies that the effect of the coupling between the centre-of-mass and internal dynamics is controlled by the temperature of the particle collection, and therefore expected to be small at low temperatures.

More importantly, comparing the eigenstates and energy eigenvalues of $\hat{H}^\text{tot}$ (\eqref{eq:CM-plane-wave} and~\eqref{eq:CM-plane-wave-energy}) with those of the single particle free Hamiltonian (\eqref{eq:plane-waves} and~\eqref{eq:plane-wave-energy}) it is clear that we can treat the centre-of-mass dynamics like those of a single particle with mass $M$, momentum $\bm{k}^\text{tot}$, and non-commutative parameter $\tilde\lambda$. With this in mind, we can revisit the classicality condition of \eqref{eq:classical-condition}, which for a collection of particles now requires
\begin{equation}\label{eq:macroscopic-classical-condition}
	r \ll \frac{4\tilde\lambda\, d^2 M E_\text{tot}}{\hbar^2}
	= \frac{4\lambda d^2 m E_\text{tot}}{\hbar^2}.
\end{equation}
Now the energy is extensive, since, by \eqref{eq:CM-plane-wave-energy}, we have that
\begin{align*}
	E_\text{tot} = N\cdot \frac{2\hbar^2}{m\lambda^2}
	\sin^2\left(\frac{\langle k\rangle\, \lambda}{2}\right)
	= N \langle E\rangle,
\end{align*}
where we have defined the average momentum and energy,
\[
	\langle k\rangle \defeq \frac{1}{N} k^\text{tot},\quad\text{and}\quad
	\langle E\rangle \defeq \frac{2\hbar^2}{m\lambda^2}
	\sin^2\left(\frac{\langle k\rangle\, \lambda}{2}\right).
\]
Finally, we can repeat the back-of-the-envelope estimate that we performed in section~\ref{sec:classical-lim} with the new condition of \eqref{eq:macroscopic-classical-condition}. We again take $\lambda$ to be of the order of a Planck length, $d$ to be on the order of $1\SI{}{\centi\meter}$, and consider a collection of electrons with average energy $1\SI{}{\electronvolt}$. Supposing we have a number of electrons on the order of Avogadro's number, we can expect to see interference suppression at distances on the order of
\begin{align*}
	r & \lesssim \frac{10^{23} (10^{-35}\, \SI{}{\meter})
		(10^{-2}\, \SI{}{\meter})^2 (10^{-31}\, \SI{}{\kilogram})
		(10^{-19}\, \SI{}{\kilogram\square\meter\per\square\second})}
	{(10^{-34}\, \SI{}{\kilogram\square\meter\per\second})^2} \\
	  & = 100\, \SI{}{\meter}.
\end{align*}
This is much more realistically detectable. Conversely, to observe suppression at distances of the order of a meter for this number of particles requires an average energy orders of magnitude less than an electron volt, making the interference suppression entirely observable at non-relativistic energies. This represents a significant improvement over the situation in the Moyal plane, where, although a quantum-to-classical transition exists, it is only observed for particles travelling at speeds well above the speed of light, even when collections of particles are considered~\cite{pittaway2021quantum}.

It is worth finally mentioning that experiments of the type outlined above seem almost within the realm of possibility. One promising approach may be splitting and recombining Bose-Einstein condensates~\cite{van2008longitudinal} in a Mach-Zehnder interferometer. Another might involve controlling and interfering a levitated nanosphere~\cite{delic2020cooling,tebbenjohanns2021quantum,millen2015cavity}. However, none of these experiments achieve control over quite the number of particles assumed in our estimates. Moreover, one should be careful in directly applying the conclusions above to these situations as there is a true environment in these experiments that will most likely lead to higher levels of decoherence and suppression of interference, since the suppression predicted above applies to an ideally isolated system.  With more control over the environment (and larger numbers of particles), these types of experiments may, however, offer a way of probing this phenomenon.

\section{Conclusion}\label{ch:conclusion}

In summary, we have treated a two-pinhole interference pattern in the fuzzy sphere formalism of NCQM, both for individual particles and collections of particles. In so doing, we have seen how the fuzzy sphere formalism not only supports a natural mechanism for quantum-to-classical transition without the need for an external heat bath, but, moreover, also addresses several key issues with a similar transition in lower-dimensional NCQM frameworks such as the Moyal plane.

Our findings are important for three reasons. Firstly, they reinforce the notion that it really is the small-scale structure of space responsible for suppression of quantum behaviour at a macroscopic scale. Secondly, they demonstrate that in three dimensions the transition is capable of being observed for realistic numbers of particles travelling at non-relativistic speeds. Thirdly, they uncover an additional system parameter involved in controlling the strength of suppression, namely the distance at which measurement is performed. The latter is an important testable prediction of our theory.

Altogether, the results are promising and wholly support the proposed link between the microscopic structure of space and macroscopic emergence of classicality.


\appendix

\section[Coherent State Transformation Laws]{Transformation Laws of Coherent States}
\label{app:su2-transformations}

In this short appendix we derive the transformation law for the action of a plane wave on a Glauber coherent state $\ket{\bm{z}}$.

As noted in section~\ref{sec:plane-waves}, the plane waves represent (all of the) $\SU2$ group elements. A general $g\in\SU2$ can be parameterised as
\[
	g \equiv g(\bm{q}) = \exp[\frac{i}{2}\bm{q}\cdot\bhat{\sigma}],
\]
by some dimensionless vector $\bm{q}$ with norm $q\in[0,2\pi]$ and direction $\bhat{q}$. This can be equivalently written (by Taylor expanding the exponential, and using $(\bhat{q}\cdot\bhat{\sigma})^2 = I$),
\begin{equation}\label{eq:expanded-su2-element}
	g(\bm{q}) = \cos(q/2) + i\sin(q/2)\,\bhat{q}\cdot\bhat{\sigma}.
\end{equation}
Thanks to the Lie algebra isomorphism $\hat{x}_i \mapsto \lambda\sigma^i$, we can represent $g$ as an operator on $\Hc$ as
\[
	\hat{\Pi}(g) = \exp[i\bm{q}\cdot\frac{\bhat{x}}{2\lambda}],
\]
which formally resembles a plane wave with wavenumber $\bm{q}$ (see \eqref{eq:plane-waves}), only $\bm{q}$ is dimensionless, and so does not represent a true wavenumber. We introduce dimensions by replacing
\(
\bm{q} \equiv 2\lambda\, \bm{k},
\)
where $\bm{k}$ is a true dimensionful wavenumber.

Note that, since $\hat\Pi(g)$ commutes with $\hat{r}^2$ (as each $\hat{x}_i$ does), it preserves the $\hat{r}^2$-eigenspaces (\textit{i.e.}~those with fixed $j$, as per \eqref{eq:jm}). In particular, plane waves preserve the vacuum state,
\(
\hat\Pi(g)\ket{0} = \ket{0},
\)
and in the $j=1/2$ irrep they act as ordinary $\SU2$ matrices,
\begin{equation}
	\hat\Pi(g)\ket{j=1/2,m} = g \begin{bmatrix}
		\ket{j=1/2,m=+1/2}\\\ket{j=1/2,m=-1/2}\end{bmatrix}.
\end{equation}
This is merely a statement of the $j=1/2$ Wigner $D$-function entries (see section 4.3.4 in~\cite{abers2003quantum}, for instance).

Now consider the action of a plane wave on a coherent state. The following computation closely mirrors \eqref{eq:rotation-of-plane-wave}, so we likewise define $\hat{J}_i\equiv\frac{1}{2\lambda}\hat{x}_i$ in this context. Then
\begin{equation}\label{eq:plane-wave-coherent-state-1}
	\hat{\Pi}(g)\ket{\bm{z}}
	= e^{-\frac{1}{2}\bar{z}_\alpha z_\alpha}\, \exp[z_\alpha
		\left(\hat\Pi(g)\,a^\dagger_\alpha\,\hat\Pi(g)^\dagger\right)] \ket{0},
\end{equation}
so the problem reduces to deriving the transformation law for the boson creation operators $a_\alpha^\dagger$ under the conjugation
\(
e^{i\bm{q}\cdot\bhat{J}} a_\alpha^\dagger e^{-i\bm{q}\cdot\bhat{J}}.
\)
But it is easily shown that these boson creation operators transform like rank-$1/2$ spherical tensor operators with respect to the $\hat{J}_i$; this just amounts to checking (see section 5.2.3 in \cite{abers2003quantum}, for instance) the commutation relations
\begin{equation}
	\begin{split}
		[\hat{J}_3,a_\alpha^\dagger] &= \left(\frac{3}{2}-\alpha\right) a_\alpha^\dagger,\\
		[\hat{J}_+,a_\alpha^\dagger] &= \delta_{\alpha 2}\,a_1^\dagger,\\
		[\hat{J}_-,a_\alpha^\dagger] &= \delta_{\alpha 1}\,a_2^\dagger,
	\end{split}
\end{equation}
where $\hat{J}_\pm\defeq\hat{J}_1\pm i\hat{J}_2$, as usual. Such tensor operators transform under conjugation with $\hat\Pi(g)$ by the $j=1/2$ Wigner $D$-matrices (see section 5.2 of~\cite{abers2003quantum}, for instance),
\[
	e^{i\bm{q}\cdot\bhat{J}} a_\alpha^\dagger e^{-i\bm{q}\cdot\bhat{J}}
	= g_{\beta\alpha}a_\beta^\dagger.
\]
Inserting this into \eqref{eq:plane-wave-coherent-state-1} reduces it to
\(
	\hat\Pi(g)\ket{\bm{z}} = \ket{g\bm{z}},
\)
a remarkably simple transformation law, reminiscent of \eqref{eq:rotation-of-plane-wave}. In summary, in terms of dimensionful wavenumber $\bm{k}$, the action of a plane wave on a coherent state is given by
\begin{equation}\label{eq:plane-wave-coherent-state}
	\begin{split}
		e^{i\bm{k}\cdot\bhat{x}}\ket{\bm{z}}
		&= \ket{g(\bm{k})\,\bm{z}}, \quad\text{where} \\
		g(\bm{k}) &\defeq e^{i\lambda\bm{k}\cdot\bhat\sigma}
		= \cos(\lambda k)+i\sin(\lambda k)\bhat{k}\cdot\bhat\sigma.
	\end{split}
\end{equation}

\section[Coherent State Matrix Elements]{Matrix Elements with Coherent States}
\label{app:coherent-state-mels}

In this appendix we develop the necessary theory to compute the matrix element of a general function $g(\hat{n})$ of the boson number operator with respect to a set of coherent states. The key step involves introducing new creation operators, then expanding the coherent states in the Fock number basis of these new operators; this step is presented in~\cite{scholtz2018classical}. The original work here comprises the special cases we consider. In particular, we derive a compact closed form for the special case where $g$ is a polynomial, and, more notably, we treat the case where $g=g_{H,l}$, the (non-commutative) spherical Hankel function. The latter is vital to the discussion in section~\ref{sec:spherical-waves}, as well as the main calculation of chapter~\ref{ch:pinhole}.

\subsection[General functions]{General functions \texorpdfstring{$g(\hat{n})$}{}}
\label{app:coherent-state-mels-general}

Consider some function $g(\hat{n})$ of the number operator $\hat{n}=a_\alpha^\dagger a_\alpha$. We wish to compute the matrix element
\(
\mel{\bm{z}^1}{g(\hat{n})}{\bm{z}^2},
\)
where the $\ket{\bm{z}^i}$ are (possibly distinct) coherent states. As in~\cite{scholtz2018classical}, we start by introducing new creation operators,
\begin{equation}\label{eq:new-creation-ops}
	A_i^\dagger\defeq\frac{1}{\sqrt{R_i}}z^i_\alpha a_\alpha^\dagger,
	\quad\text{where}\quad R_i=\bar{z}^i_\alpha z^i_\alpha,
\end{equation}
and then rewriting the coherent states in terms of the the new operators,
\[
	\ket{\bm{z}^i} = \hat{D}_i\ket{0} \equiv e^{-R_i/2+\sqrt{R_i}A_i^\dagger}\ket{0}.
\]
Next, we expand the exponential in the above displacement operator to express each coherent state in the basis of Fock states (those corresponding to the new creation operators) as
\begin{equation}\label{eq:coherent-in-Fock-basis}
	\ket{\bm{z}^i} = e^{-R_i/2}\sum_{n=0}^\infty
	\frac{R_i^{n/2}}{\sqrt{n!}}\ket{n}_i,
\end{equation}
where the subscript on the ket indicates the boson mode (\textit{i.e.}~distinguishes between application of each of the two $A_i^\dagger$). Substituting this expansion into the desired matrix element then gives
\begin{equation}\label{eq:coherent-mel-unsimplified}
	\mel{\bm{z}^1}{g(\hat{n})}{\bm{z}^2}
	= e^{-(R_1+R_2)/2}\sum_{n,m}\frac{R_1^{n/2}R_2^{m/2}}{\sqrt{n!\, m!}}g(m)
	\prescript{}{1\hspace{-1.0pt}}{\ip{n}{m}}_2
\end{equation}
Now clearly \(\prescript{}{1\hspace{-1.0pt}}{\ip{n}{m}}_2 = 0\) whenever $n\neq m$, since each $\ket{n}_i$ can be written (by binomial expansion on $(A_i^\dagger)^n$) as a linear combination of number states of the form $\ket{l_1,l_2}$, where $l_1+l_2=n$ --- to wit, $\ket{n}_i$ indeed contains $n$ total particles. The overlap $\prescript{}{1\hspace{-1.0pt}}{\ip{n}{n}}_2$ is less trivial,
\begin{align*}
	\prescript{}{1\hspace{-1.0pt}}{\ip{n}{n}}_2
	= \left[A_1,A_2^\dagger\right]^n
	= \left(\frac{1}{\sqrt{R_1R_2}}\bar{z}_\alpha^1 z_\alpha^2\right)^n,
\end{align*}
which is easily derived using the Leibniz rule. Finally, inserting the relevant Fock state overlaps, \eqref{eq:coherent-mel-unsimplified} reduces to
\begin{equation}\label{eq:coherent-mel}
	\begin{split}
		\mel{\bm{z}^1}{g(\hat{n})}{\bm{z}^2}
		&= e^{-(R_1+R_2)/2}\sum_{n}g(n)
		\frac{\left(\bar{z}_\alpha^1 z_\alpha^2\right)^n}{n!}.
	\end{split}
\end{equation}
Deriving a closed form for the sum of this series is often straightforward using the umbral calculus~\cite{roman2005umbral}. That said, this little-known calculus is thankfully seldom needed; indeed, we proceed to explicitly compute the sum for polynomial functions $g$ without reference to umbral calculus.

\subsection[Polynomials]{Polynomials \texorpdfstring{$g(\hat{n})$}{}}
\label{app:coherent-state-mels-polynomials}

Since sides of \eqref{eq:coherent-mel} are linear in $g$, we may focus without loss of generality on the case $g(\hat{n})=\hat{n}^k$. For convenience, we also define $K\defeq \bar{z}_\alpha^1 z_\alpha^2$. Then we may rewrite $g(n)$ using the well-known relationship~\cite{roman2005umbral}
\[
	g(n) = n^k = \sum_{m=0}^k\stirling{k}{m}n^{\underline{m}},
\]
where $n^{\underline{m}}$ denotes the falling factorial,
\[
	n^{\underline{m}} \defeq \prod_{k=0}^{m-1}(x-k),
\]
and $\stirling{k}{m}$ are the Stirling numbers of the second kind. Then the sum in \eqref{eq:coherent-mel} becomes
\begin{align*}
	\sum_{n=0}^\infty g(n)\frac{K^n}{n!}
	 & = \sum_{m=0}^k\stirling{k}{m} \sum_{n=0}^\infty
	n^{\underline{m}} \,\frac{K^n}{n!}                 \\
	 & = \sum_{m=0}^k\stirling{k}{m} \sum_{n=0}^\infty
	(n+m)^{\underline{m}} \,\frac{K^{n+m}}{(n+m)!}     \\
	 & = \sum_{m=0}^k\stirling{k}{m}K^m
	\cdot \sum_{n=0}^\infty	\frac{K^n}{n!}              \\
	 & = e^K T_k(K).
\end{align*}
We may shift the sum on the second line, since $n^{\underline{m}}=0$ for all $n\in\{0,1,\dots,m-1\}$. The functions $T_k(x)\defeq \sum_{m=0}^k\stirling{k}{m}x^m$ are the Touchard polynomials, which are easily looked up, say in the OEIS~\cite{oeisA106800}. We tabulate the first few Touchard polynomials in table \ref{tab:touchard}.
\begin{table}[t]
	\centering
	\begin{tabular}{l|l}
		$\bm{k}$ & $\bm{T_k(x)}$     \\ \hline
		$0$      & $1$               \\
		$1$      & $x$               \\
		$2$      & $x^2+x$           \\
		$3$      & $x^3+3x^2+x$      \\
		$4$      & $x^4+6x^3+7x^2+x$
	\end{tabular}
	\caption{Touchard polynomials $T_k(x)$ up to $k=4$}
	\label{tab:touchard}
\end{table}
In summary, we have the special case of \eqref{eq:coherent-mel},
\begin{equation}\label{eq:coherent-mel-poly}
	\mel{\bm{z}^1}{\hat{n}^k}{\bm{z}^2} = e^{-(R_1+R_2)/2+K}T_k(K),
\end{equation}
for $R_i = \bar{z}_\alpha^iz_\alpha^i$ and $K = \bar{z}_\alpha^1 z_\alpha^2$. This of course recovers the usual overlap of coherent states in the case $k=0$.

\subsection[Hankel function]{Spherical Hankel function
	\texorpdfstring{$g(\hat{n}) = g_{H,l}(\hat{n},k)$}{}}
\label{app:coherent-state-mels-hankel}

In this subsection, we consider one more special case --- we compute the (leading-order behaviour of the) matrix element of the (asymptotically-expanded) non-commutative spherical Hankel function (see \eqref{eq:asymptotic-Hankel}),
\[
	g(\hat{n}) \equiv g_{H,l}(\hat{n},k)
	= \frac{e^{i(\hat{n} + l + 1)\kappa}}{(i\hat{n})^{l + 1}},
\]
with respect to an identical pair of coherent states,
\(
\ket{\bm{z}^1} = \ket{\bm{z}^2} \equiv \ket{\bm{z}}.
\)
This particular case, while highly specific, is of great importance to the discussion in section~\ref{sec:spherical-waves} and the subsequent calculation in chapter~\ref{ch:pinhole}, so it warrants discussion.

Now, we can of course apply the result of section~\ref{app:coherent-state-mels-general} in the present context, but some care is needed --- the coherent state $\ket{\bm{z}}$ is a superposition of all boson number states, including the vacuum state, so the singular $1/\hat{n}$ causes divergence in a na\"ive calculation. This is not really a problem: not only is the overlap
\(
\abs{\ip{0}{\bm{z}}}^2 = e^{-R/2}
\)
vanishingly small for $R \equiv \bar{z}_\alpha z_\alpha \gg 1$ (which is assumed for the asymptotic form of $g_{H,l}$ as given above), but we can entirely remove the divergence by defining $g$ piecewise as the regular solution $g_{J,l}$ within some region around the origin and the irregular solution $g_{H,l}$ outside of this region. Equivalently, we may as well use only the irregular form, but exclude the point $n=0$ from our computations. As such, we invoke \eqref{eq:coherent-mel}, but begin the sum at $n=1$, whereby
\begin{align*}
	H_{l,\kappa}(R)
	 & \equiv \ev{\frac{e^{i(\hat{n} + l + 1)\kappa}}{(i\hat{n})^{l + 1}}}{\bm{z}} \\
	 & = e^{-R} \sum_{n=1}^\infty \frac{e^{i(n+l+1)\kappa}}{n!\, (in)^{l+1}} R^n   \\
	 & = \frac{e^{i(l+1)\kappa}}{i^{l+1}} e^{-R} (e^{i\kappa} R)
	\sum_{n=0}^\infty \frac{(e^{i\kappa} R)^n}{(n+1)!\, (n+1)^{l+1}},
\end{align*}
using the shorthand notation $H_{l,\kappa}(R)$ for the desired matrix element. Now the remaining sum can be written as a generalised hypergeometric function,
\[
	{\prescript{}{p}{F}}_{\!\!q}\left[\begin{matrix}
			a_1, a_2, \cdots a_p \\
			b_1, b_2, \cdots b_q
		\end{matrix};\, e^{i\kappa} R\right].
\]
Indeed, if $\beta_n$ denotes the $n$th coefficient,
\[
	\beta_n \defeq \frac{1}{(n+1)! (n+1)^{l+1}},
\]
then we have $\beta_0 = 1$ (as required by convention of the coefficients in the series of a hypergeometric function), and the ratio
\[
	\frac{\beta_{n+1}}{\beta_n} = \frac{(n+1)^{l+2}}{(n+2)^{l+2}(n+1)},
\]
is clearly a rational function of $n$, from which we can read off the values $p=q=l+2$,
$a_1 = a_2 = \cdots = a_p = 1$, and $b_1 = b_2 = \cdots = b_q = 2$, giving
\begin{equation}\label{eq:coherent-mel-hankel}
	H_{l,\kappa}(R) = \frac{e^{i(l + 2) \kappa}}{i^{l+1}} R e^{-R}\,
	{\prescript{}{l+2}{F}}_{\!\!l+2}\left[\begin{matrix}
			1, 1, \cdots 1 \\
			2, 2, \cdots 2
		\end{matrix};\, e^{i\kappa} R\right].
\end{equation}
It is worth noting that (by the ratio test) the series representing this hypergeometric function is both convergent on all of $\mathbb{C}$ and entire, since $p = l + 2 = q$~\cite{volkmer2014note}.

There is no algebraic representation of ${\prescript{}{l+2}{F}}_{\!\!l+2}$, so \eqref{eq:coherent-mel-hankel} is the best we can do without further approximation. However, since we are already using an asymptotic form of $g_{H,l}$, we should similarly seek only the leading-order large-$R$ asymptotic behaviour of $H_{l,k}$. Thankfully, the relevant hypergeometric function has a well-known asymptotic expansion, given in equations 1.2 and 1.3 of~\cite{volkmer2014note}, which in our context simplifies to
\begin{equation*}
	{\prescript{}{l+2}{F}}_{\!\!l+2} \left[\begin{matrix}
			1, 1, \cdots 1 \\
			2, 2, \cdots 2
		\end{matrix};\, z\right] \sim e^z z^{-(l+2)} \sum_{k=0}^\infty c_k\, z^{-k},
\end{equation*}
for $z\to\infty$. The coefficients $c_k$ are defined recursively; for our purposes it suffices to merely note that $c_0 = 1$, since it is clearly the $k = 0$ term that defines the leading-order behaviour in this asymptotic limit. As such, we simply truncate the series after the first term. Substituting the appropriate argument and simplifying, we arrive at the leading-order large-$R$ behaviour of our matrix element,
\[
	H_{l,\kappa}(R) \sim \frac{e^{R(\cos\kappa - 1) + iR\sin\kappa}}{(iR)^{l+1}}.
\]

\bibliographystyle{unsrtnat}
\bibliography{references}
\end{document}